\documentclass[letterpaper,twocolumn,10pt]{article}
\PassOptionsToPackage{hyphens}{url}
\usepackage{usenix2019_v3}
\usepackage{tikz}
\usepackage{cite}
\usepackage{amsmath,amssymb,amsfonts}
\usepackage[ruled,linesnumbered]{algorithm2e}
\usepackage{listings}
\usepackage{pifont}
\usepackage{graphicx}
\usepackage{textcomp}
\usepackage{xcolor}
\usepackage{subcaption}
\usepackage{booktabs}
\usepackage{array}
\usepackage{multirow}
\usepackage{tabularx}
\usepackage{rotating}
\usepackage{lscape}

\usepackage[subtle]{savetrees}  
\usepackage{titlesec}
\titlespacing*{\section}
{0pt}{2ex}{1ex}
\titlespacing*{\subsection}
{0pt}{1ex}{1ex}

\newcommand{\sysname}[0]{P\textsuperscript{2}IM\xspace}
\newcommand{\sysnamefull}[0]{Processor-Peripheral Interface Modeling\xspace}
\newcommand{\propname}[0]{P\textsuperscript{2}IE\xspace}
\newcommand{\propnamefull}[0]{Processor-Peripheral Interface Equivalence\xspace}

\newcommand{\creg}[0]{{\tt CR}\xspace}
\newcommand{\dreg}[0]{{\tt DR}\xspace}
\newcommand{\sreg}[0]{{\tt SR}\xspace}
\newcommand{\csreg}[0]{{\tt C\&SR}\xspace}

\newcommand{\point}[1]{\vspace{0.8ex}\par\noindent{\bf #1:}}

\newcommand*{\rom}[1]{\textit{\expandafter\romannumeral #1}}

\newcolumntype{L}[1]{>{\raggedright\let\newline\\\arraybackslash\hspace{0pt}}m{#1}}
\newcolumntype{C}[1]{>{\centering\let\newline\\\arraybackslash\hspace{0pt}}m{#1}}
\newcolumntype{R}[1]{>{\raggedleft\let\newline\\\arraybackslash\hspace{0pt}}m{#1}}

\begin{document}
\date{}

\title{\Large \bf $\bf P^2IM$: Scalable and Hardware-independent Firmware Testing via \\Automatic Peripheral Interface Modeling (extended version)}

\author{
{\rm Bo Feng}\\
Northeastern University\\
feng.bo@husky.neu.edu
\and
{\rm Alejandro Mera}\\
Northeastern University\\
mera.a@husky.neu.edu
\and
{\rm Long Lu}\\
Northeastern University\\
l.lu@northeastern.edu
} %

\maketitle

\begin{abstract}
    \small
    Dynamic testing or fuzzing of embedded firmware is severely limited by
    hardware-dependence and poor scalability, partly contributing to the
    widespread vulnerable IoT devices. 
    We propose a software framework that continuously executes a given firmware
    binary while channeling inputs from an off-the-shelf fuzzer, enabling
    hardware-independent and scalable firmware testing.
    Our framework, using a novel technique called \sysname, abstracts diverse
    peripherals and handles firmware I/O on the fly based on automatically
    generated models. \sysname is oblivious to peripheral designs and generic to
    firmware implementations, and therefore, applicable to a wide range of
    embedded devices. 
    We evaluated our framework using 70 sample firmware and 10 firmware from real
    devices, including a drone, a robot, and a PLC. It successfully executed
    79\% of the sample firmware without any manual assistance. We also performed a
    limited fuzzing test on the real firmware, which unveiled 7 unique unknown
    bugs.

\end{abstract}
     
\definecolor{codegreen}{rgb}{0,0.6,0}
\definecolor{codegray}{rgb}{0.5,0.5,0.5}
\definecolor{codepurple}{rgb}{0.58,0,0.82}

\lstdefinestyle{mystyle}{
    commentstyle=\color{codegreen},
    keywordstyle=\color{blue},
    numberstyle=\color{codegray},
    stringstyle=\color{codepurple},
    basicstyle=\ttfamily\footnotesize,
    breakatwhitespace=false,         
    breaklines=true,                 
    captionpos=b,                    
    keepspaces=true,                 
    numbers=left,                    
    numbersep=5pt,                  
    showspaces=false,                
    showstringspaces=false,
    showtabs=false,                  
    tabsize=2,
    frame=single,
    xleftmargin=2em,
    framexleftmargin=1.5em,
}

\graphicspath{{fig/}}

\lstset{style=mystyle}

\section{Introduction}
\label{sec:intro}

Microcontrollers, or MCU, are commonly used for building IoT (Internet of
Things) and modern embedded devices, thanks to their high energy-efficiency,
extensible connectivity, and adequate computing power. As MCU devices become
widely deployed in various scenarios, ranging from smart homes to industrial
systems, their security has been raised as a major concern among users and
operators. As demonstrated in recent reports~\cite{papp2015embedded}, software
vulnerabilities cause the majority of attacks on MCU devices, resulting in not
only digital but also physical damages.

MCU firmware (i.e., the whole software stack on MCU) contains vulnerabilities
just as computer software does. Most MCU vulnerabilities are virtually the same in
nature as their computer counterparts. Therefore, it would be ideal if the
proven vulnerability discovering techniques on computers, such as fuzz-testing
or fuzzing, can be applied to MCU firmware. However, in reality, off-the-shelf
fuzzers cannot directly test firmware, which partly contributed to the fact that
many firmware is not sufficiently tested for security vulnerabilities
\cite{not_well_tested}.

The inapplicability of fuzzers on MCU boils down to the lack of a platform where
firmware can execute while taking inputs from fuzzers. Existing emulators cannot
help because none of them emulates the whole range of MCU peripherals (i.e.,
unable to run firmware). Some recent works addressed this issue using
hybrid emulation~\cite{avatar,surrogates,charm,inception}, which forwards peripheral operations to real
devices. Although this approach creates a platform to run firmware with a fuzzer,
the platform is fairly slow and can hardly scale due to the hardware dependence.

\begin{figure}[t]
    \centering
    \includegraphics[width=7cm]{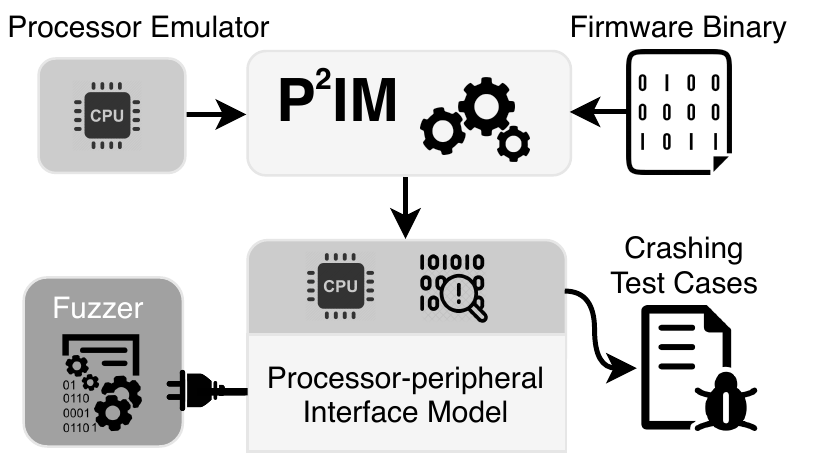}
    \caption{Framework Overview}
    \label{fig:overview}
\end{figure}

We present a novel approach to firmware fuzzing. We design a framework to run
and test MCU firmware at scale without any hardware dependence. Our framework
takes a firmware binary as input and hosts an unmodified fuzzer (AFL \cite{afl}) as a
drop-in component. Using a generic processor emulator (QEMU), the framework
executes the firmware and handles its peripheral accesses while channeling the
fuzzing input and feedback between the firmware and the fuzzer. Figure
\ref{fig:overview} shows an overview of the framework.

The key technique used in our framework is called \sysname (or \sysnamefull). It
automatically models the I/O behaviors of a wide range of peripherals while
treating peripherals themselves as black boxes. The generated models satisfy a
property we formulated, called \propnamefull. We show that when this property is
satisfied, an emulated execution of firmware can continue smoothly (e.g.,  no
crash, stall or skipped peripheral operations) without requiring any dependent peripherals (either real or
emulated).

We evaluated our framework using 70 sample firmware and 10 firmware from real
MCU-based devices, including a drone, a robot, and a PLC (Programmable Logic
Controller). The result shows that our framework can continuously run 79\% of
the sample firmware without any crash, stall or skipped peripheral operations. We also
performed basic fuzzing (without memory sanitizer) on the real firmware and
discovered 7 unique and previously unknown bugs.

\section{Roadmap \& Overview}
\label{sec:overview}

\subsection{MCU Firmware \& Testing}

Firmware generally means any low-level software that controls hardware in a
computing device. In this paper, we focus on firmware for microcontrollers
(MCU). These devices are cost- and power-effective computers built for specific
purposes, such as the motion controller of a self-balancing robot, the engine
control units (ECU) in a car, etc. STM32L010F4~\cite{STM32L010F4} is an example
of ultra-low-power MCUs commonly used in IoT devices. It carries an ARM
Cortex-M0+ processor at 32MHz, 16KB flash as the persistent storage, 2KB of RAM,
and a wide range of peripherals.

MCU firmware is usually a monolithic piece of software that contains peripheral
device drivers, a tiny OS or system library, and a set of specialized logics or
applications. For example, the firmware in a MCU-based drone contains the
drivers for all onboard peripherals, either a small real-time operating system
or vendor-customized system library, and the PID (proportional, integral and
derivative) controller among other application-level logics.  We note that MCU
vendors rarely use general-purpose OS such as Linux to build MCU firmware. Due
to hardware constraints, they prefer an OS specifically designed for MCU or
simply use a thin system library in lieu of a stand-alone OS (called bare-metal
devices). In the rest of the paper, we refer to MCU firmware simply as firmware
for brevity.

Due to the fast development and wide adoption of MCU devices in cyber-physical
and IoT systems, the security issues of these devices, often caused by
vulnerable firmware~\cite{papp2015embedded}, have led to severe consequences and
become a major concern among users and operators~\cite{hacked_camera}. 
To
mitigate vulnerable MCU devices, researchers recently proposed techniques for
fuzz-testing firmware~\cite{avatar,surrogates,inception}. These techniques allow
partial execution of firmware on an emulator while forwarding unsupported
operations (e.g., peripheral I/O) to real hardware.

This line of work allowed fuzzing to be applied to firmware. But due to the
hardware dependence and slow forwarding, fuzzing through these partial emulators
can hardly scale up. For instance, the number of parallel fuzzing runs is
limited by the availability and capacity of the dependent hardware; the speed of
each fuzzing run is severely capped by the I/O forwarding, which is three orders
of magnitude slower than native I/O~\cite{surrogates}. As a result, high scalability, the key
requirement for effective software fuzzing, cannot be achieved when using
partial emulation that depends on slow and limited hardware.

\subsection{Open Challenges}

If the state-of-the-art fuzzers could work directly on firmware at scale, the
significant values of these fuzzers demonstrated on computer software (e.g.,
unparalleled vulnerability discovery ability) can automatically transfer to MCU
firmware, which can tremendously help reduce vulnerabilities and improve
security of MCU devices. However, despite the previous efforts aiming at this
goal~\cite{avatar,surrogates,charm,inception}, we still identified the following
open challenges that prevent fuzzers for computer software from being effective
on firmware. 

\point{Hardware Dependence}
Previous efforts on firmware fuzzing require certain hardware (e.g.,
peripherals). This is due to incomplete hardware emulation. Moreover, such
dependent hardware is much slower than emulators running on computers. As a
result, the hardware dependence introduces orders of magnitudes of delays to
fuzzer execution.
Moreover, hardware dependence also critically limits parallelism. For instance,
one dependent peripheral can only be used by one fuzzing session. Therefore,
highly parallel fuzzing, which is the key to fuzzers' success on computer
software, is not achievable.  

\point{Wide Range of Peripherals}
Due to the poor performance and scalability caused by hardware dependence, some
recent work proposed purely emulation-based fuzzing of firmware. In fact,
fuzzing software on a fully emulated platform has been found useful for a long
time in cases where software under test cannot be instrumented or is only
available in binary forms. However, creating fully emulated MCU has proven
impractical and no existing emulators offer generic MCU support. This is mainly
because of the highly heterogeneous MCU hardware in general and the wide range
of peripherals in particular. Each firmware may interact with a distinct set of
peripherals, which can be customized by the MCU vendor. Peripherals of the same
type but different models/brands often have different specifications and
interfaces. Therefore, a specially customized emulator is often required for
fuzzing or testing a new firmware. Building such emulators remains a manual
task, which is not only error-prone but impossible to catch up with the large
and fast-increasing number of MCU devices.

\point{Diverse OS/System Designs}
In addition to the hardware-related challenges unique to MCU, the software also
poses challenges to firmware fuzzing that are currently unaddressed. Unlike
general-purpose computers, whose OSes are dominated by a few mainstream options
that follow similar designs, MCU devices use a much larger and more diverse set
of OSes that are significantly different from each other. Many MCU devices do not
even have a typical OS but a system library that manages hardware and task
scheduling. The diverse OS/system designs among MCU means that OS-specific
fuzzing methods, which existing system fuzzers use (e.g., syscall fuzzers), are
not applicable to firmware. In other words, generic firmware fuzzing should be
OS-agnostic and not make assumptions about the OS/system designs. 

\point{Incompatible Fuzzing Interfaces}
Another software-related challenge unique to firmware fuzzing is about the
interfaces through which fuzzer-generated inputs are channeled into firmware
execution. For computer software fuzzing, the input interfaces are well-defined
and uniform (e.g., via files or standard I/O). However, firmware reads all
inputs via peripherals, which come in many different types and have their own
access conventions. Making the matter more complicated, different drivers in
firmware may configure the same peripheral differently and then perform I/O
through different interfaces. As a result, the input interfaces supported by
existing fuzzers are incompatible with firmware. Moreover, manually adding
support for every peripheral I/O interface to fuzzers can be a daunting task, if
possible at all.

\vspace{1em}\noindent We note that the aforementioned challenges are unique to
MCU firmware fuzzing. There are other open problems facing software fuzzing in
general, such as better input generation, more effective error detection,
etc. However, this work is focused on tackling the challenges unique to firmware
fuzzing. We consider improving general fuzzing techniques orthogonal and
out-of-scope for this paper.  

\subsection{Our Approach}

We present a novel approach to MCU firmware fuzzing, which overcomes the
challenges discussed before. We design a framework that supports fuzzers as
drop-in components to test firmware in a scalable and hardware-independent
fashion. The framework aims to solve the MCU-imposed fuzzing challenges while
allowing fuzzers to focus on performing and improving their own job (i.e.,
generating inputs and finding bugs). The goal of our framework is to bridge the
wide open gap between fuzzers and firmware. It allows existing fuzzers to test
firmware without any knowledge about the software and hardware design of MCU. It
also facilitates the development of specialized fuzzers for firmware. 

Our approach is novel in that it neither relies on any hardware nor emulates
peripherals. We introduce a form of {\em approximate MCU emulation} for
supporting firmware testing and fuzzing. More importantly, we provide a method to
automatically generate approximate emulators based on firmware binaries.
The approach is inspired by our observation that firmware can execute on an
emulator without real or fully emulated peripherals, as long as the emulator
provides the firmware with {\em acceptable} inputs from peripherals when needed.
Such inputs do not have to be the same as what a real peripheral would produce.
But they do need to pass firmware's internal checks to avoid disrupting firmware
execution. For a given firmware, our approximate emulator uses a generic
processor/ISA emulator (e.g., one for ARM Cortex-M) and a model, automatically
built for the firmware, that captures what constitutes an acceptable input for
each peripheral accessed by the firmware. We find that this kind of approximate
emulation can comprehensively exercise a firmware (i.e., covering most firmware
code), and therefore, is sufficient for supporting fuzzing and other types of
firmware analyses that examine the control or data behaviors of firmware (as
opposed to testing functional correctness), such as taint analysis, invariant
detection, etc. 

Next, we provide the necessary background on peripheral interfaces. We then
define a property that an approximate MCU emulation must meet in order to be
acceptable for supporting firmware fuzzing. At the end of the section, we
discuss the high-level design of our framework that enables the
approximate firmware execution and supports firmware fuzzing.

\subsection{Processor-Peripheral Interfaces}

\begin{figure}[t]
    \centering
    \includegraphics[width=0.9\columnwidth]{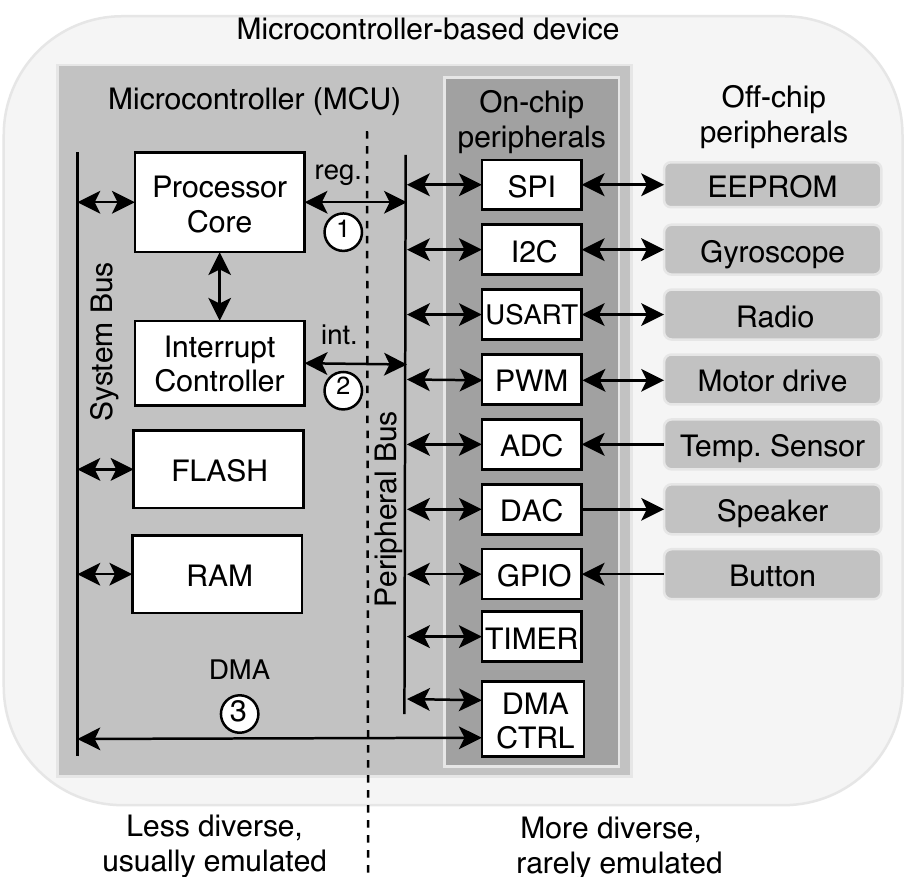}
    \caption{Architecture diagram of MCU devices. 
    Firmware running on processor core interacts with peripherals via \textcircled{\small{1}} memory-mapped registers, \textcircled{\small{2}} interrupts, and \textcircled{\small{3}} DMA. 
    }
    \label{fig:dev_arch}
\end{figure}

Peripherals are indispensable from MCU devices and of great varieties. As shown
in Figure~\ref{fig:dev_arch}, they can be on-chip or off-chip. On-chip
peripherals typically serve as the proxy through which data travels between
firmware and off-chip peripherals. Some on-chip peripherals are not externally
connected and provide simple functionalities needed by firmware (e.g., timers).
In this paper, we only consider on-chip peripherals (or peripherals for short)
because firmware cannot access off-chip peripherals directly. As illustrated in
Figure~\ref{fig:dev_arch}, there are three types of peripheral I/O interfaces
exposed to firmware, namely memory-mapped registers, interrupts, and direct
memory access (DMA). Firmware performs all I/O through these interfaces. We
refer to the first two types (\textcircled{\small{1}} and
\textcircled{\small{2}} in Figure~\ref{fig:dev_arch}) as {\em
processor-peripheral interfaces} as they connect processors and peripherals. 

We note that this work covers the processor-peripheral interfaces and not DMA.
We leave DMA out of scope for this paper because it is extremely difficult to
model automatically and its I/O behavior is heavily dependent on internal
designs of individual peripherals, which our method is oblivious of to be
generic. Nonetheless, DMA is not frequently used by MCU peripherals, which tend
to exchange small amounts of data with firmware 
(only 2 out of 70 firmware tested in \S\ref{sec:unittest} use DMA).

We define a new property, related to peripheral I/O modeling, for 
MCU emulators. 
It is called \propnamefull (or \propname). Satisfying this property
means that: (1) the emulator emulates the {\em processor-peripheral interfaces},
rather than peripherals themselves used by the firmware, and (2) the emulated
interfaces are equivalent to those of the peripherals expected by the firmware,
in terms of their impact to firmware execution. The formulation of \propname is
based on our experience with firmware analysis for a wide range of MCUs.
We observed that providing equivalent processor-peripheral
interfaces is sufficient for a generic emulator, without any peripheral
emulation, to comprehensively execute and test/fuzz firmware. A
\propname-enabled emulator handles peripheral I/O operations by providing the
processor-peripheral interfaces and mimicking their external behaviors. 

We also define an empirical test for \propname: the property is satisfied if the
firmware running on the peripheral-agnostic emulator never {\em crashes, stalls,
or skips operations} due to peripheral I/O errors. A {\em crash} may happen when
the firmware tries to read/write data from/to a peripheral but encounters a
fatal error, such as illegal memory access or unsupported peripheral operations.
A {\em stall} may occur when the firmware waits for a peripheral state to change
but the emulator fails to recognize and handle it. Under a similar situation,
the firmware may eventually give up on waiting and {\em skip operations},
causing parts of firmware code to be unreachable. For instance, before reading
data from a peripheral such as ADC (analog-to-digital converter), firmware needs
to wait for a memory-mapped register bit to be set, which indicates data is
ready. If the emulator fails to set such bits, firmware stalls without showing
any signs of errors. Alternatively, after a long wait, the firmware simply skips
the operations (not only the input operation but also subsequent operations
depending on the input). 

We use the definition of \propname to guide our framework design. We use
the empirical test of \propname as a way to verify if our framework generates
emulators that satisfy this property. 

By focusing on the interface equivalence
(i.e., generalizable), rather than emulating every peripheral (i.e.,
non-generalizable), we demonstrate that it is {\em possible to automatically
build approximate emulators} for MCU devices equipped with a wide range of
peripherals. This automated generation of MCU emulators is the key to
hardware-independent, scalable, and high-coverage firmware testing.

\subsection{Framework Overview}

The framework, for the first time, allows firmware to be dynamically
tested and fuzzed without using any MCU devices, hardware peripherals, or human
assistance.

\begin{figure}[t]
    \centering
    \includegraphics[width=\columnwidth]{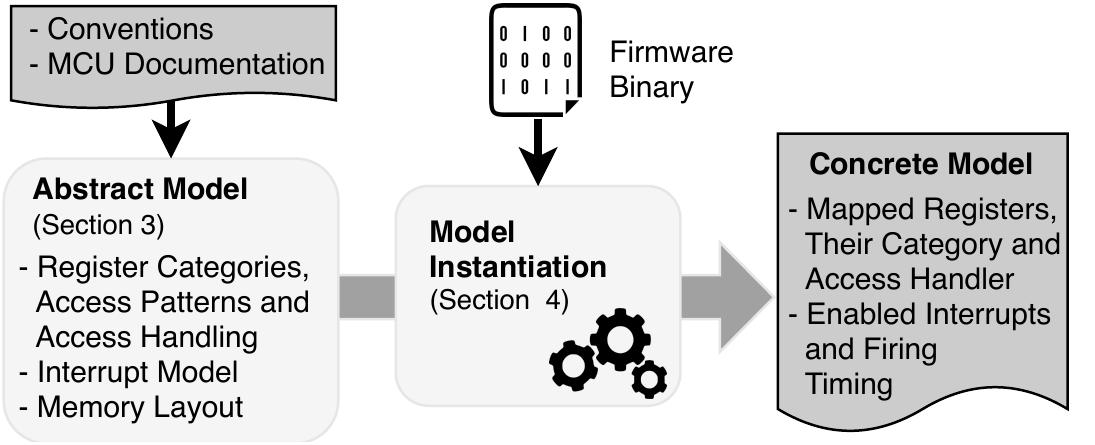}
    \caption{\sysname workflow}
    \label{fig:workflow}
\end{figure}

The model derivation process, called \sysnamefull (or \sysname), contains two
steps, as shown in Figure~\ref{fig:workflow}. First, an abstract model is
defined for a broad class of MCU architectures (e.g., ARM Cortex-M). An abstract
model captures the generic patterns and conventions that firmware follows and acceptable input when
accessing processor-peripheral interfaces. Such information is readily available
in MCU device datasheets or processor documentation. An abstract model also
contains a customizable interrupt firing strategy suitable for the entire MCU architecture
class. Defining an abstract model is a manual and offline process done by domain
experts. It is practical because it only happens once for each class of MCU
architectures (only a few architecture classes are common among MCU) and, using
our template, defining an abstract model for a new architecture class does not
require too much effort. Abstract models do not vary much across different
architecture classes. We discuss the definition of the abstract model for ARM
Cortex-M, serving as a template for other MCU architectures, in \S\ref{sec:def}.

The second step is model instantiation, which is fully automatic and needed for
every firmware to be tested. It instantiates the abstract model defined for the
MCU architecture of a given firmware. It concretizes the abstract model with the
firmware-specific information, such as where specifically each peripheral
register is mapped in memory and what interdependency among the registers, if
broken, may impact firmware execution. This firmware-specific (or
device-specific) information is necessary due to the high heterogeneity of MCU
(e.g., devices using the same architecture often have different peripheral and
interface specifications). Without this information, emulators cannot provide
the processor-peripheral interfaces equivalent to the real ones (i.e., achieving
\propname). Our framework automatically infers the firmware-specific information
using a technique called explorative firmware executions. An instantiated model
tells the emulator what constitutes access to peripheral interfaces and how
such access should be handled based on its type and the runtime condition. We
discuss the details about model instantiation in \S\ref{sec:inst}.

Besides \sysname, another important part of our framework design is the
support of fuzzers as drop-in components and the feeding of fuzzing inputs to
the firmware execution. Our framework does not have special requirements for
fuzzers, which are simply treated as black-box input generators. The framework
channels the fuzzing input into the peripheral interface access handlers in the
emulator, which feed the fuzzing input when the firmware expects raw input data
(as opposed metadata or status input) from peripherals. Our framework also
provides standard coverage feedback to fuzzers, collected through the emulator.
We discuss the fuzzer setup and the fuzzing results on real firmware in
\S\ref{sec:fuzzingcase}
\section{Abstract Model Definition}
\label{sec:def}

As illustrated in Figure~\ref{fig:workflow}, 
the first step of modeling the processor-peripheral interfaces is to build an
abstract model for a target MCU architecture class. This is the only manual step
in \sysname and should be fairly straightforward for embedded system engineers
or security researchers with basic knowledge about MCU firmware and hardware. An abstract
model captures the generic patterns and conventions that firmware follows when
accessing the processor-peripheral interfaces. For instance, firmware for MCU
devices based on ARM Cortex-M typically access different types of peripheral
registers via memory-mapped I/O (i.e., peripheral registers are mapped to a
fixed region in memory for firmware to read/write). Firmware for these devices
also enables a range of peripheral interrupts mainly for performing asynchronous
I/O. 

We define an abstract model for ARM Cortex-M, the most popular architecture
class for IoT devices, which can be used as a template for building abstract
models for other MCU architectures (see \S\ref{sec:discuss} for more details). 
The model generalizes peripheral registers
into four types and provides the access patterns and handling strategy for each type (i.e.,
how an emulator should identify each type of registers and handle each access to a peripheral register based on its
type). This access pattern-based register type identification and
type-based register access handling is generically applicable to all
peripherals on Cortex-M. Therefore, emulators can perform them without requiring
any knowledge about specific peripherals (e.g., what kind of peripheral does a
register belong to) or knowledge about peripheral internal designs. The model
abstracts peripheral interrupt firing as a special input channel and allows
customizable interrupt firing strategies. Also included in the abstract model
are the locations of the basic memory segments, such as the RAM, the flash, the
mapped register region, which remain the same for devices using the same MCU
architecture and are specified in MCU documentations\footnote{The flash region,
where firmware is loaded, may vary on some devices. This information is
available in device datasheet.}. 
On ARM Cortex-M MCUs, peripheral registers are mapped to 
{\tt 0x40000000}--{\tt 0x5fffffff} memory segment as required by the architecture design \cite{v6m_arch_ref_man} 
\cite{v7m_arch_ref_man} (i.e., this is an architectural requirement that all hardware and software using this architecture need to follow). 
\sysname considers each memory word in this segment a
potential memory-mapped register.

\subsection[]{Register Category, Access Patterns and Handling}
\label{sec:regcat}

\point{Control Registers}
Control registers of peripherals, or \creg, are used mostly by firmware to
control or configure peripherals. For example, firmware sets the corresponding
bits in USART's \creg to enable the transmitter or interrupts or to set the baud
rate. %
\textbf{\creg Access pattern}: We observed a read-modify-write (RMW)
pattern unique to \creg, whereby firmware first reads a \creg, then modifies
the configuration parameters in it, and finally writes the value back to the
register. Firmware follows the RMW pattern when accessing \creg because
firmware can only write at word/register granularity and the RMW pattern
avoids inadvertently changing other parameters co-exist in the same register.
\sysname uses the RMW pattern to identify \creg. 
In some rare case (e.g., a \creg contains only one configuration parameter),
firmware may write directly to it without following the RMW pattern. In this
case \sysname, due to the write-on-first-access pattern of \dreg defined below,
\sysname can mis-categorize such a \creg into \dreg. However, in most cases,
this mis-categorized register is never read afterward (i.e., for one-time
peripheral configuration). Therefore, this kind of register mis-categorization
does not impact firmware execution or needs correction. 

\textbf{\creg Access handling}: %
Once a peripheral is configured by the firmware, the peripheral operates accordingly and 
rarely changes value of \creg (i.e., peripheral configuration). Therefore, \sysname
models each \creg as a {\em non-volatile} memory word. When firmware reads a
\creg, the emulator returns the value previously written to the \creg. If a
\creg is read without being explicitly written before, which can happen after a
hardware reset, the emulator simply returns zero, which is the
default value for \creg in most cases. 

\point{Status Registers}
\label{sec:mf_rc_sr}
A status register, or \sreg, is a set of flags (i.e., each flag may contain 1 or
more bits) that indicate the internal states of a peripheral. 
During runtime, peripherals update their \sreg as their status change
(peripherals can also use interrupts to notify firmware of status changes, which
is discussed in \S\ref{sec:interruptmod}). Before performing certain peripheral
I/O operations, firmware polls the corresponding \sreg bits to make sure the
peripheral is ready. For example, firmware reads data from USART only when the
data-reception flag in a \sreg is set, indicating some data has been received.
Otherwise, firmware simply waits. In many cases, if the necessary
\sreg flags are not set, firmware ceases to boot or stalls infinitely (e.g., the
system-clock-ready flag in a clock manager peripheral). On the other hand,
setting the wrong \sreg bits can cause firmware to crash. For example, certain
\sreg bits being set means fatal peripheral errors, which can switch firmware
into recovery or debug mode that requires external intervention. Therefore,
properly handling firmware access to \sreg is critical for achieving \propname
and undisrupted execution and testing of firmware.

\textbf{\sreg Access pattern}:
These registers are used by firmware to check peripheral states. \sysname
categorizes a newly discovered register as \sreg if the first access to the
register is an unconditional read and the read value is later evaluated in a
condition.  
For some \sreg, the first access on them can be a write, for example, when
firmware acknowledging a peripheral error. In this case, \sysname could
initially mis-categorize the \sreg into \dreg due to the second \dreg access
pattern defined below. However, \sysname automatically corrects such mistakes
later by leveraging our observation that firmware often reads \sreg
continuously (i.e., to poll peripheral state under synchronous I/O) but not
other types of registers. Therefore, when firmware continuously polls on a
previously categorized \dreg, \sysname adjusts its category to \sreg and locks
its type. We call it polling pattern of \sreg. 

\textbf{\sreg Access handling}: Since \sysname does not model peripherals
themselves and is oblivious to their internal designs, its handling of \sreg is
not based on knowing the semantics of the flags or the registers. Instead, the
\sreg handling aims to dynamically infer an acceptable register value at each
\sreg read so that firmware can continue executing (i.e., the value can pass the
firmware's internal checks and lead to subsequent peripheral I/O operations
guarded by this \sreg). \sysname uses a technique called {\em explorative
execution} to automatically infer acceptable \sreg values during runtime. This
technique belongs to the firmware-specific part of \sysname (i.e., the model
instantiation part), which is discussed in \S\ref{sec:inst}. On the other hand,
handling \sreg write is much simpler and the same for all firmware. \sysname
treats \sreg writes as no-ops, which are ignored by the emulator. This is because
\sreg are volatile and values written by firmware only matter to peripherals
internally and are not read back by firmware. In other words, \sreg write is
one-way and the value is transient and does not affect firmware execution.
\propname can be achieved without handling \sreg writes.

\point{Data Registers}
\label{sec:mf_rc_dr}
Data registers, or \dreg, are the main channel through which raw data flows from
peripherals to firmware. Oftentimes, data read by firmware through \dreg
originates from off-chip peripherals (e.g., Zigbee radio)
or a remotely connected device (e.g., a supervisory computer in SCADA system).
For example, SPI peripheral holds the data it received from an off-chip peripheral (e.g., Zigbee radio) in its
\dreg, which is then read by firmware as input. Data also flows in the opposite
direction. Firmware writes output data in the \dreg and then SPI sends it to
an off-chip peripheral. 

\textbf{\dreg Access pattern}: Firmware only reads a \dreg (i.e., taking raw
input from a peripheral) after confirming that the peripheral is in a ready
state by checking the corresponding \sreg. Based on this unique access pattern
of \dreg, \sysname categorizes a newly discovered register as \dreg if reading
the register is preceded by an \sreg read and conditional on a flag in the
\sreg. Sometimes firmware writes to \dreg directly without checking any \sreg. 
\sysname uses this write-on-first-access as another access pattern for
identifying \dreg. 

\textbf{\dreg Access handling}: \dreg of all peripherals collectively dominate the
inputs to firmware, and therefore, they are ideal \textbf{{\em fuzzing interfaces}}. Our
framework uses modeled \dreg to feed fuzzing and testing inputs during runtime.
These inputs are generated by a drop-in fuzzer, which may or may not be aware of
firmware/peripheral specifics (our current prototype uses unmodified AFL). Upon
each \dreg read, the emulator returns the next word from the fuzzing input as the
register value. For other types of dynamic analysis, the input source can be
replaced with, for example, previously recorded inputs (for bug/execution
reproduction) or specially crafted inputs (for taint analysis). Similar to \sreg
writes, \sysname ignores \dreg writes for the same reason that they do not
affect firmware execution.

\point{Control-Status Register}
\label{sec:mf_rc_csr}
A control-status register, or \csreg, is a hybrid register whose bits are split
between two purposes: control/configuration bits (same to \creg) and status bits
(same to \sreg). Although hybrid registers allow for higher utilization of
register bits, they greatly complicate both peripheral hardware and firmware
designs. Therefore, they are not commonly used in modern MCU devices, which have
abundant memory address space for mapping peripheral registers. In practice, we
only observed some rare use of \csreg. We have not seen other types of hybrid
registers, such as control-data or status-data registers, which are
theoretically possible but impractical. 
\textbf{\csreg Access pattern}: \creg bits of \csreg are modified in the RMW
pattern during the peripheral configuration phase. \sreg bits of \csreg are
accessed during the peripheral operation phase. The configuration phase
proceeds the operation phase. They do not overlap.
As a result, \sysname often categorizes \csreg as \creg in the first
place. However, such inaccurately categorized registers are corrected later
when \sysname observes the \sreg access pattern on them.

\textbf{\csreg Access handling}: 
For each \csreg access, firmware operates on either
the \creg bits or the \sreg bits, but not both because they are used at
different stages during firmware execution. Since handling \sreg bit access
requires firmware-specific information (similar to handling \sreg register
access), \csreg handling is not covered by the abstract model but by the
instantiated model, which is discussed in \S\ref{sec:inst}.

\point{Remarks} Although the register access patterns and the type
identification method are purely empirical, we find that in practice they work
fairly reliably and accurately across a wide range of peripheral devices (see
\S\ref{sec:eval} for the evaluation results). We attribute this practical and
promising results to two factors: (1) the register types we defined are
generically applicable to all peripherals; (2) the type-based access patterns
were observed and generalized from a variety of real MCU devices; (3) trade-offs
are carefully made when designing the type-based access patterns. Specifically, we 
observed the write-on-first-access pattern not only on \dreg, but also on \creg
and \sreg in some occasions. We still use this pattern to 
identify \dreg despite that certain \creg and \sreg might be mis-categorized. 
This trade-off is made and justified by the following considerations. 
First, this pattern is most commonly seen on \dreg.  By using this pattern for detecting
\dreg, we can achieve the best overall register categorization accuracy. Second, a 
\sreg mis-categorized by this pattern is often corrected later on by the polling 
pattern unique to \sreg (i.e., this mis-categorization is temporary). Third, a \creg mis-categorized by this pattern (e.g., a \creg
contains only one configuration parameter) does not impact firmware execution/testing or needs correction because firmware generally does not read or take input from \creg. 

\subsection{Interrupt Firing}
\label{sec:interruptmod}

Apart from the register categories and handling, the abstract model also defines
how emulators should fire necessary interrupts on behalf of peripherals in order
to satisfy \propname and support continuous firmware execution or testing. 

In essence, interrupts are a special type of inputs to firmware. They notify
firmware of certain hardware events and trigger the corresponding interrupt
service routines (ISR), which are interrupt handlers implemented by peripheral
drivers in firmware. For instance, an interrupt may signal the firmware that
input data is ready. Then the corresponding ISR is invoked and reads the input
data from a \dreg. 

A processor emulator, such as QEMU, often includes a virtual interrupt
controller, which could dispatch fired interrupts to software. However, since these
emulators do not emulate MCU peripherals, no peripheral interrupt is fired when
using them to run a firmware, despite that the firmware may crash or stall for
other reasons before it gets ready for servicing interrupts.   

\sysname abstractively models interrupts as a sequence of timing-based inputs,
with each input corresponding to an enabled interrupt. When such an input comes
in, the emulator generates and dispatches the matching interrupt to the
firmware. The emulator detects what interrupts are enabled by the firmware
during runtime (discussed in \S\ref{sec:intruntime}). \sysname allows both the
sequence and the timing of interrupts to be customized based on different
fuzzing strategies (e.g., purely random generation, mutation from crafted seeds,
etc.). Our current prototype uses a
simple interrupt firing strategy: enabled interrupts are fired in a round-robin
fashion at a fixed interval (e.g., after every 1,000 basic blocks executed). The
interval is defined using the number of executed basic blocks, rather than
absolute time (e.g.,  clock ticks). This basic block-based interval definition
supports arbitrary timings to be specified for emulators to fire interrupts.
More importantly, using basic block counts to measure interrupt intervals allows
for deterministic replay of interrupt sequences, and therefore, yields
reproducible fuzzing/testing results. This reducibility is required for fuzzing,
but usually hard to achieve when using existing fuzzers with asynchronous
interrupts enabled. 

We note that defining more advanced interrupt firing strategies is the job of
drop-in fuzzer or fuzzer operators and is out of the scope for \sysname.
Although the current interrupt firing strategy is simple, it already leads to a
very high firmware code coverage as shown in the evaluation section. 
\subsection[]{Infeasible Peripheral Inputs \& False Positives}
\label{sec:fp}
Under the current abstract model definition, \sysname can trigger code paths in 
firmware that are infeasible on real devices. This is because hardware
peripherals may only generate certain inputs and fire interrupt at certain
patterns, whereas \sysname allows fuzzers to generate random peripheral inputs
or adopt arbitrary interrupt timing. Such infeasible inputs and code paths may
cause false positives during fuzzing (i.e., a crash/hang is caused by an
infeasible input/path, rather than by a firmware bug). However, as a generic
firmware testing framework, \sysname does not prune potentially infeasible
inputs or code paths. Instead, \sysname leaves the task of input pruning, which
is part of the input generation process, to the testing tools running on top of
\sysname (e.g., a fuzzer). This design decision is made for two reasons. First,
input generation and input quality control are among the core tasks of fuzzers
and other dynamic testing tools. \sysname is designed to support these tools and
not in the position to interfere with these tasks. Second, as observed in
\cite{periscope}, peripherals such as Wi-Fi radio are vulnerable to remote
attacks (e.g., an attacker sending malformed network packet), and once
compromised, can generate unexpected input or interrupt timing. Therefore,
testing firmware with unexpected/infeasible peripheral inputs might be desirable
in some cases. 

Nevertheless, we did not see in our extensive experiments any crashes/hangs
that were caused by the infeasible inputs or code paths introduced by \sysname, despite
that no input pruning was performed. All crashes/hangs detected on \sysname were 
reproducible on real hardware (i.e., they were caused by true bugs in tested firmware), except for two false hangs that were caused by \sysname. We analyze these two cases and the limitations of \sysname that caused them in \S\ref{sec:miscat}.  %
\section{Automatic Model Instantiation}
\label{sec:inst}

As illustrated in Figure~\ref{fig:workflow}, 
the second step of \sysname is the instantiation of an abstract model defined in
the first step, producing a full model for a given firmware. The instantiated
model guides the emulator to identify and handle I/O operations through the
process-peripheral interfaces. During the instantiation step, the
firmware-specific information needed for providing \propname is added to an
abstract model. 

The model instantiation process is fully automatic and uses the explorative
execution technique. The instantiation is on-demand and interleaved with the
firmware fuzzing/testing process. 
The fuzzing process invokes the model instantiation process when it encounters
unmodeled or unhandled peripheral access. The model instantiation process 
terminates and the fuzzing resumes when the model becomes stable and no 
new information (e.g., newly identified registers) is added to the model for a while. Note that 
this switch is not triggered by the model reaching a certain level of precision.
We call each invocation of the model 
instantiation process ``one round of model instantiation''. Multiple rounds of model instantiation can happen at different points throughout a firmware fuzzing/testing process. The model 
instantiation process is deterministic and repeatable. An instantiated 
model can be reused for the same firmware. 
The process relies on a customized QEMU that
emulates the Cortex-M instruction set and a generic interrupt controller but not
MCU peripherals. 

Specifically, an instantiated model contains the following automatically
inferred firmware-specific information, which concretizes the abstract model:
(1) identified memory-mapped registers, their memory locations, and types; (2)
the access handling strategies for each type of registers, or each use site when
needed; (3) the enabled interrupts and the firing strategy. An instantiated 
model does not contain any information about peripheral configurations or internals.

\begin{figure}[t]
    \centering
    \includegraphics[width=0.95\columnwidth]{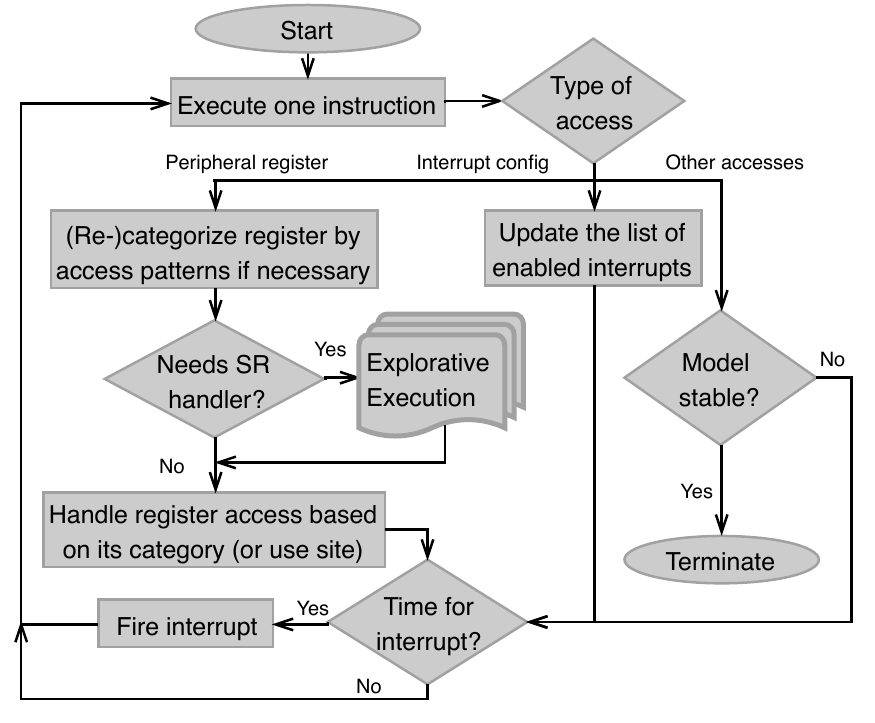}
    \caption{Model instantiation workflow}
    \label{fig:mi}
\end{figure}

Figure~\ref{fig:mi} shows the high-level workflow of the model instantiation process. 
It executes the firmware on a generic processor emulator which does not emulate any peripherals. 
It continuously instantiates the model when firmware accesses the processor-peripheral interfaces. 
Upon a register access, it (re-)categorizes the register if necessary 
as described in \S \ref{sec:mi_ri}.  
Then it handles the register access as described in \S \ref{sec:reghand}. 
Specially, for an SR read, it checks whether a handler exists. 
If not, it performs the explorative execution to automatically generate a handler (\S \ref{sec:reghand}). 
It also monitors interrupt configurations by the firmware and fires interrupts
when needed (\S \ref{sec:intruntime})

\subsection{Register Identification}
\label{sec:mi_ri}

The goal of register identification is to detect the memory-mapped registers
exposed by peripherals and determine their types according to our category and access pattern definition
(\S\ref{sec:regcat}). It identifies and categorizes all registers that are accessed by the 
firmware as it runs on \sysname. 
During the model instantiation process, \sysname monitors
firmware's access to the memory segment reserved for peripheral registers (e.g.,
{\tt 0x40000000}--{\tt 0x5fffffff} on ARM Cortex-M MCUs \cite{v6m_arch_ref_man}\cite{v7m_arch_ref_man}, as captured in the
abstract model). \sysname considers each accessed memory word in this segment a
memory-mapped register. 

Although detecting such registers is straightforward, determining their types is
fairly challenging because \sysname or the emulator does not have any knowledge
about the semantics of the registers or the peripherals that the registers
belong to. Overcoming this challenge, \sysname determines the type of a newly
identified register based on the per-type register access patterns that we
empirically observed and generalized from a large set of MCU peripherals (\S\ref{sec:regcat}). %

\point{Peripheral Association}
In addition to identifying peripheral registers and their types, \sysname also
groups them based on if  they belong to the same peripheral. This grouping is
needed for accurately handling certain \sreg accesses (discussed in
\S\ref{sec:reghand}). It only considers peripheral association and is unaware of
peripheral types or characteristics. \sysname identifies the groups based on the
spatial adjacency and alignment of registers' memory addresses.

\subsection{Register Access Handling \& Explorative Execution}
\label{sec:reghand}

\sysname provides strategies for type-based register access handling, which
instructs the emulator,  upon register access, what actions it should take,
including return what value to the firmware or what internal state to update.
The strategies for handling \dreg and \creg accesses are straightforward and
uniform across firmware. They are part of the abstract model defined in
\S\ref{sec:regcat}. 

On the other hand, the strategies for handling \sreg, including the \sreg bits
in \csreg, are much more complicated. The strategies may vary across different
\sreg as well as different use sites of the same \sreg. For example, the most
significant bit in two different \sreg, $SR_1$ and $SR_2$, have completely
different meanings. At one point of firmware execution, setting the bit in
$SR_1$ is needed for the firmware execution to continue without stalling but
setting it in $SR_2$ crashes the firmware. At different points of firmware
execution, setting the bit in $SR_1$ causes the opposite. Therefore, strategies
for handling \sreg accesses need to consider firmware specifics, individual
registers, and their access contexts. As part of the model instantiation step,
\sysname automatically derives \sreg handling strategies using a technique
called {\em explorative execution}, which is the focus of this subsection.

The high-level idea of the explorative execution is as follows. When the
firmware execution encounters a new \sreg access site, \sysname pauses and
snapshots the execution and starts the explorative execution. By spawning
multiple parallel worker threads, the explorative execution concurrently
searches for the best value for the \sreg, resumes the original firmware
execution, and returns the \sreg value to the firmware. The key challenges
addressed by our design of this technique include: constructing a tractable
search space of candidate \sreg values; determining the scope of the explorative
execution (or the termination condition for the workers); defining what
qualifies the best \sreg value; reducing the frequency of explorative
executions. We describe our solutions to these challenges below.

\point{Search Space Construction}
The search space could be intuitively constructed by including all possible
values for an \sreg. This  interprets to $2^{32}$ candidate values on a 32-bit
MCU and in turn requires the explorative execution to spawn $2^{32}$ worker
threads to test the candidate values, which obviously is infeasible. Instead, we
construct a much smaller and tractable search space by taking advantage of the
fact that bits/flags in \sreg are usually independent and only a single flag is
checked at a time. 
Our search space contains only 32+1
candidate values, each with a single bit set in an \sreg plus a zero (all bits
clear). The explorative execution spawns one worker thread to test each
candidate value. In each thread, the candidate value is returned to the firmware
as the value of the \sreg. All worker threads execute in parallel. \sysname
monitors their progress and picks a winner (i.e., the thread with best candidate
value) at the end of the explorative execution.

\point{Termination of Explorative Workers}
When the explorative execution should terminate a worker thread is another
design question. If too early, the worker thread may have not reached the use of
the \sreg that is critical to firmware execution. If too late, the explorative
execution becomes too long and can cause significant delay or even halt the
firmware execution (e.g., due to encountering another \sreg read whose access
handling strategy has not been derived yet). We experimented several termination
conditions and different life spans of work threads. We found one that works
well in practice and keeps the runtime overhead low. 
It terminates a worker thread when it is about to return to the next level
callee (i.e., when the current call stack frame, where the \sreg read happened,
is popped). The rationale is that firmware usually reads an \sreg in the same
function where it decides if further I/O operations can be performed based on
the \sreg value. Therefore, having the explorative execution continue beyond
function boundaries does not yield additional benefits for finding the best \sreg
value, despite that it significantly complicates the thread monitoring mechanism
and slows down \sysname. It is worth noting that many worker threads exit before
they reach the termination point because the assigned \sreg values are
unacceptable to the firmware.

\point{Qualified Workers and \sreg Values}
After all worker threads terminate, \sysname determines which threads or
candidate \sreg values qualify for potentially advancing firmware execution. It
then picks the best among the qualified values to return to the original
firmware execution, which concludes the explorative execution. The qualification
criteria are: (1) the thread did not crash or stall; (2) if all threads crashed
or stalled, choose those caused by other factors than the current \sreg (i.e.,
the crash/halt site is not dependent on the \sreg value). 
Among the qualified worker threads and the candidate \sreg values they
represent, \sysname selects the best based on the number of \dreg accesses,
guarded by the \sreg, that were observed during thread execution. When
multiple equally good \sreg values are found, \sysname randomly picks one as the
best value (and records the choice to make the model instantiation process 
deterministic and repeatable). The design of the worker qualification and 
selection aligns with the
definition of \propname: input values that are acceptable to firmware and unlock
meaningful operations are used in place of inputs from real peripherals to
achieve \propname and sufficient for supporting firmware fuzzing/testing.

\point{Minimizing Explorative Executions via \sreg Grouping} 
The design of the explorative execution discussed so far treats individual \sreg
accesses independently. If implemented as is, this design can cause frequent
explorative executions (upon every \sreg read during firmware execution) and
thus high overhead. But on the other hand, access handling strategies need to be
derived for every use of every \sreg as explained earlier. We address these two
conflicting needs by optimizing our design via \sreg grouping. The idea is that
an access handling strategy derived for one \sreg at one location, though not
universally applicable to all \sreg, can be reused for the same \sreg accessed
in similar locations. Specifically, we group \sreg accesses based on their
context, defined by a four-tuple $(r, cs, bbl, conf)$, where $r$ is the \sreg;
$cs$ is the signature of the call stack at the time of the \sreg access; $bbl$
is the ID of the basic block in which the SR read occurred; $conf$ is the
peripheral configuration hash trivially generated from \creg values at the time of the
\sreg access, which does not contain any semantic information for peripheral  
configurations (such as whether the receiver is on or off). The configuration 
hash is included because different \creg values
cause the firmware to check \sreg differently. For example, firmware only checks the
data-reception flag of USART when the receiver is enabled via \creg. With \sreg
grouping, similar \sreg accesses can reuse the same handling strategy, which
increasingly reduces the frequency of explorative executions as \sysname
instantiates the model.

\subsection{Interrupt Identification}
\label{sec:intruntime}
Another task that \sysname performs during the model instantiation step is
collecting the firmware-specific information about interrupts. MCU architectures
(e.g., Cortex-M) support hundreds of different interrupts. But a particular
device or its firmware may only use a small subset of supported interrupts.
Moreover, during runtime, firmware sometimes dynamically disables and re-enables
interrupts as needed. If an emulator fires an unused or disabled interrupt,
firmware execution can stall or crash because firmware commonly uses a simple
dead loop as the default handler for unused interrupts.

\sysname maintains a list of currently enabled interrupts during firmware
execution. It taps into the virtual interrupt controller of QEMU (Nested
Vectored Interrupt Controller, NVIC), which the firmware configures to
enabled/disable interrupts. Drawing from the list of enabled interrupts, the
interrupt firing strategy defined as part of the abstract model (\S\ref{sec:interruptmod})
decides when to fire what interrupts.  

\subsection{\sysname Implementation}
\label{sec:impl}
We implemented our framework using QEMU as the base processor emulator (without
any peripheral emulation capability). 
Our implementation includes 2,202 lines of C code added to QEMU (mostly for dynamic firmware execution instrumentation), 173 lines of C code for fuzzer integration, and 
1,199 lines of Python code for the explorative execution part of \sysname. We use AFL as the drop-in fuzzer in our current prototype, which has no built-in support or awareness of MCU firmware.

We implemented register categorization, peripheral identification, type-based
register access handling, and SR read grouping logic inside two QEMU functions,
namely \texttt{unassigned\_mem\_read} and \texttt{unassigned\_mem\_write},
where accesses to memory-mapped peripheral registers are directed to. 
For fast prototyping, we implemented the complex logic of explorative execution
using Python. But the worker threads still run natively on QEMU. 
The interrupt identification and firing logic are implemented based on the QEMU's
virtual interrupt controller (NVIC). The logic monitors firmware's accesses to
\texttt{NVIC\_ISERx} and \texttt{NVIC\_ICERx} registers to detect enabled interrupts.
It fires interrupts via the
\texttt{armv7m\_nvic\_set\_pending} interface exposed by NVIC. 
AFL's emulation mode (used for fuzzing un-instrumented binaries) only supports
user-mode emulation, which is incompatible with firmware
emulation~\cite{qemuom}. TriforceAFL~\cite{triforceafl} builds a bridge for AFL
to be connected to the full system emulation mode of QEMU. We used TriforceAFL's
code when implementing the fuzzer integration part of our framework, which
allows fuzzers to be dropped in without modifications. 
During runtime, the fuzzer integration code channels inputs generated by the
fuzzer to firmware execution through \dreg accesses. It collects code coverage
information via the QEMU instrumentation and returns the information to the
fuzzer. 

\section{Evaluation \& Fuzzing Results}
\label{sec:eval}

We evaluated our framework from three different angles: (1) whether it satisfies
\propname when executing firmware for different MCU with different OSes; (2) how
its runtime performance is in practice; (3) whether it can perform fuzz-testing
on real firmware in a fully emulated fashion (i.e., no hardware dependence) and find previously unknown bugs. 

To that end, we performed functional unit tests based on commonly used MCU
peripherals and different MCU OSes (\S\ref{sec:unittest}). We also conducted an
end-to-end test on real firmware (\S\ref{sec:firmwaretest}). Finally, using our
framework, we performed fuzzing on real firmware, found bugs, and gained
interesting insights (\S\ref{sec:fuzzingcase}). We conducted all experiments on
a moderate-spec computer with a dual-core Intel\textregistered \
Core\texttrademark \ i5-7260U CPU @ 2.20GHz, 8 GB of RAM, and Ubuntu 16.04. 

\subsection{Unit Tests on MCU Peripherals \& OSes }
\label{sec:unittest}

We designed and performed this experiment to verify if our framework can indeed
provide \propname when fuzzing firmware that: (1) access a range of peripherals
(i.e., \sysname provides generic peripheral support), (2) are designed for different
MCU SoCs (i.e., \sysname is applicable to a broad class of MCU),
and (3) use different OS/system libraries (i.e., \sysname is OS
agnostic). 
For this purpose, we collected a set of example firmware as unit test cases for 
this experiment. 

\point{Experiment Setup}
We identified the 8 most popular MCU peripherals---implemented as 
on-chip peripherals---by analyzing the entire MCU product line (1686 MCU parts) 
offered by Microchip Technology, a top global MCU vendor. We
selected these peripherals (the left column in Table~\ref{t-functional-test})
for our unit tests. We also selected 3 widely used MCU OS/system libraries,
(NuttX, RIOT, and Arduino) and 3 target MCU SoCs (STM32 F103RB, NXP MK64FN1M0VLL12,
and Atmel SAM3X8E). We selected these SoCs because they are part of the reference designs
provided by major MCU vendors, and had been integrated into automotive/marine \cite{product_DUE}, 
consumer \cite{product_F103} and 
healthcare \cite{product_K64F} products.

\begin{table}
    \centering
    \caption{Selected peripherals \& functional operations}
    \label{t-functional-test}
    \begin{tabular}[t!]{@{}lll@{}} \\ \toprule
    \multicolumn{1}{l}{\textbf{Peripheral}} & \multicolumn{1}{l}{\textbf{Functional Operations}} \\ \midrule
    \textbf{SPI} &\begin{tabular}{@{}l@{}}Receive a byte\\Transmit a byte \end{tabular}\\ \hline
    \textbf{USART} &\begin{tabular}{@{}l@{}}Receive a byte \\Transmit a byte\end{tabular}\\ \hline
    \textbf{I2C} &\begin{tabular}{@{}l@{}}Read a byte from a slave \\Write a byte to a slave\end{tabular}\\  \hline
    \textbf{GPIO} &\begin{tabular}{@{}l@{}l@{}}Read status of a pin\\Set/Clear a pin\\ Execute callback after pin interrupt \end{tabular}\\ \hline
    \textbf{ADC}  &Read an analog-to-digital conversion \\ \hline
    \textbf{DAC}  &Write a value for digital-to-analog conversion  \\ \hline
    \textbf{TIMER} &\begin{tabular}{@{}l@{}}Execute callback after interrupt\\Read counter value\end{tabular}\\ \hline
    \textbf{PWM}  &Configure PWM as an autonomous peripheral   \\ \hline
    \end{tabular}
\end{table}

We collected 70 different example firmware or test cases, each representing a unique 
and feasible combination of a peripheral, an OS, and a SoC. After booting, these
firmware simply perform the basic peripheral operations defined in
Table~\ref{t-functional-test}. We made sure these firmware run smoothly on their
target SoC. We then run them using our framework and collect the statistics and
results, including the accuracy of the instantiated model and end results.

\point{Experiment Results}
We collected the statistics on peripherals and registers accessed during the
tests, as shown in Table~\ref{t-StatisticsEvaluation}. This shows that a single
peripheral operation often incurs multiple accesses to related or dependent
peripherals of different kinds. For example, the minimum number of peripherals
accessed for during the {\tt GPIO} test is 3 and the maximum number is 15 for
{\tt I2C}. Additionally, multiple register accesses are associated with a single
peripheral operation. For instance, the minimum number of involved registers for
a {\tt GPIO} operation is 9 and the maximum number is 68 for {\tt ADC}. These
statistics show that even a simple peripheral operation can involve a complex
chain of other peripherals and many registers, highlighting the value of
\sysname and the need for automatic modeling and handling of MCU peripherals. 

\begin{table}
    \centering
    \caption{Peripherals and registers accessed during unit tests}
    \label{t-StatisticsEvaluation}
    \begin{tabular}{@{}lllllll@{}} \\ \toprule
    \multicolumn{1}{l}{\textbf{}} &\multicolumn{3}{l}{\textbf{Peripherals accessed}} & \multicolumn{3}{l}{\textbf{Registers accessed}} \\ \midrule
    \textbf{Peripheral}   &\textbf{Max.}  &\textbf{Min.}  &\textbf{Avg.} &\textbf{Max.}  &\textbf{Min.}  &\textbf{Avg.}    \\
    \textbf{I2C}    &15	 &5	 &9.0	 &54	 &18	  &35.6   \\
    \textbf{ADC}    &14	 &6	 &8.8	 &68	 &30	  &46.0   \\
    \textbf{PWM}    &14	 &7	 &10.2	 &62	 &25	  &43.2   \\
    \textbf{TIMER}  &14	 &7	 &9.7	 &47	 &26	  &38.0   \\
    \textbf{GPIO}   &13	 &3	 &7.7	 &57	 &9	      &34.3   \\
    \textbf{SPI}    &13	 &6	 &8.3	 &66	 &19	  &36.8   \\
    \textbf{USART}  &13	 &4	 &7.5	 &53	 &15	  &30.0   \\
    \textbf{DAC}    &11	 &8	 &9.5	 &60	 &35	  &47.5   \\
    \hline
    \end{tabular}
\end{table}

\begin{figure*}[t]
    \centering
    \includegraphics[width=17.5cm]{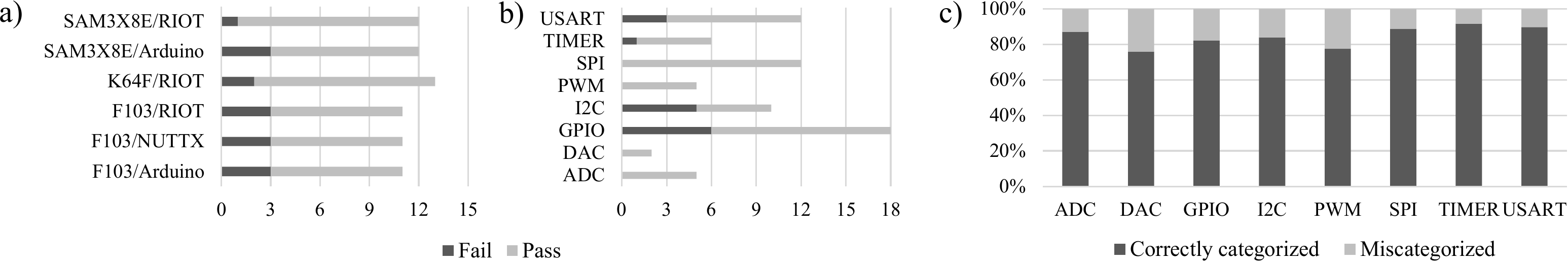}
    \caption{Unit test results aggregated by MCU SoC/OS (a) and Peripheral (b). Accuracy of register categorization (c) }
    \label{fig:UnitTestMerged}
    \end{figure*}

We also measured the accuracy of register identification and categorization.
We first manually extracted the ground truth from the MCU datasheets and then 
compared the register categorization output from our system with the ground truth.
Figure~\ref{fig:UnitTestMerged} c) shows the result aggregated by peripherals, ranging
from 76\% to 92\% (i.e., 24\% to 8\% identified registers are mis-categorized). 
There are no particular peripherals on which \sysname performs
much better or worse than others. This suggests that \sysname's accuracy does
not vary much across different types of peripherals. It also echos that \sysname is
oblivious to peripheral types or internals. We discuss the reasons of register 
mis-categorizations and their impact on firmware execution in \S \ref{sec:miscat}.

The unit test result is that 79\% (55) of the tests passed (i.e., \propname was
satisfied) while 21\% (15) failed. For a test to qualify as pass, the firmware
under test needs to properly boot, configure the peripherals, and conduct the
functionality, without any crash, stall, or operation skipping. The pass rate of
79\% may not seem very high at first glance. But considering it represents an
improvement from 0\% (i.e., no previous work can generically and automatically
model MCU peripheral I/O), we argue the result is in fact significant. The
per-MCU and per-peripheral breakdowns are shown in
Figures~\ref{fig:UnitTestMerged} a) and b). The former shows
that \sysname performs equally well across different MCU SoC and OS
combinations, suggesting it is MCU- and OS-agnostic.  The latter reveals that
\sysname encountered  failures on {\tt USART}, {\tt TIMER}, {\tt I2C}, and {\tt GPIO} but not
the other peripherals.

We found 2 general causes for failed tests. First, register 
mis-categorizations can happen when peripheral drivers fail to follow the
correct/common register access patterns. In our experiments, we observed the 
majority of failures on {\tt GPIO} and  {\tt I2C} peripherals are due to this reason. 
Second, some peripherals multiplex individual interrupts,
which can cause \sysname to fire incorrect interrupts.

Overall, this experiment shows that \sysname works reasonably well on a large
set of example firmware (using real drivers and system libraries and no
accommodation to \sysname). It allows most of the firmware to execute, without
any crash, stall, or operation skipping, on an emulator that does not support
MCU peripherals. Moreover, \sysname is shown to be agnostic to firmware's target
MCU and OS.  

\subsection{End-to-end Tests against Real Firmware}
\label{sec:firmwaretest}

This experiment examines the performance of our framework when running and
testing real MCU firmware of different kinds. These firmware are much bigger and
more complex than the unit test firmware used in the previous experiment,
although the unit tests are larger in quantity and more diverse in terms of the
used peripherals and target MCU and OSes.  This and the previous experiment
together show \sysname's ability of handling diverse and complex firmware. 

\point{Experiment Setup}
We selected 10 firmware of real MCU-based devices used for different purposes,
ranging from drones to industrial control systems. 
They are full-fledged firmware and contain all the common firmware components, including 
the kernel (e.g., scheduler, interrupt handler, system libraries), drivers, 
console, application logic, etc. 
They collectively cover 4 MCU models (from 3 top MCU vendors by revenue
\cite{mcu_vendor_rank}), 4 OSes and a diverse set of peripherals (Table 
\ref{t-case-study-firmware}). 
Moreover, the underlying SoC used in these
firmware are often used in other embedded or IoT devices. 
We evaluated both 
the model instantiation mechanism (in the current section) and the fuzzing performance 
(\S \ref{sec:fuzzingcase}) on the 10 selected firmware. 
The details about the 10 firmware are presented in Table \ref{t-case-study-firmware} in Appendix \ref{sec:fm_info}. 
A brief description for each firmware is as follows: 

{\em Self-balancing Robot}: 
This is the motion controller firmware in a robot architecturally similar to
commercial personal transporters (e.g., Segway PT). Even basic vulnerabilities in such
firmware, such as integer overflows, can lead to disastrous consequences
\cite{Ariane5} or life-threatening accidents. 

{\em PLC (Programmable Logic Controller)}: PLC is a rugged embedded device for
controlling critical processes in industrial environments (e.g., assembly lines).
This selected firmware is part of a sterilizer machine and manages a PLC's communication with remote SCADA
(Supervisory control and data acquisition) systems via Modbus, an industrial
communication protocol~\cite{ModbusSpec}. Vulnerabilities in
PLC firmware are often critical as demonstrated by the Stuxnet
attack~\cite{Stuxnet}.  

{\em Gateway}:  This firmware is for a gateway device that uses the Firmata
\cite{firmata} protocol to communicate with its host computer, allowing the host
computer to access and configure MCU peripherals dynamically, such as sensors
and actuators. Security vulnerabilities in such firmware can be exploited to
remotely hijack/abuse embedded devices. 

{\em Drone}: This firmware drives the MCU-based autopilot controller in a
quad-copter similar to the Pluto Drone \cite{PlutoDrone}. It controls multiple
sensors, radios, motors, etc. and implements a suite of control algorithms, such
as PID (proportional, integral and derivative). Vulnerabilities in drone
firmware can be exploited to manipulate drones and cause physical damage.

{\em CNC}: This firmware is a Cortex-M port of the widely used Grbl milling
controller \cite{cnc_grbl}. Grbl has been used in many commercial and
open-source 3D printers, laser cutters, hole drillers, etc. This firmware
includes a G-code interpreter, the linear/circular interpolation
algorithms, and the stepper-motor control routines. Vulnerabilities in the
G-code interpreter or control routines can lead to physical injuries of machine
operators or destruction of the milling equipment. 

{\em Reflow Oven}: This firmware is for a commercial-grade reflow oven
\cite{reflow_oven} controller used for assembly of printed circuit boards (PCB).
This controller implements push-buttons and LCD as user interface, thermocouple
input, acoustic alarm, and dual output for the heating element and oven fan. The
temperature profile of controller is based on the multi-ramp \cite{ramp_pid} PID
control loop. Vulnerabilities in this firmware can compromise the industrial
processes and the quality of PCB assembly.

{\em Console}: This firmware implements all the standard utilities of the RIOT OS
 and exposes the shell through a serial console. The shell of RIOT
 implements a small but powerful interface to execute user-defined callbacks and
 other system utilities for control and diagnostic purposes. Vulnerabilities in the shell
 can compromise internal data structures of the OS and even expose a device to
 remote code execution.

{\em Steering Control}: This firmware implements the algorithm of a
steer-by-wire \cite{steer_by_wire} controller deployed in a lab-grade
self-driving vehicle. It takes commands from the main on-board computer and
translates them to electrical signals to control servomotors \cite{servomotors}.
Bugs in this or similar type of devices are the causes of multiple
documented deadly car accidents, plaints and recalls from major automotive
companies \cite{toyota_bug}. 

{\em Soldering Iron}: This firmware is an open-source version of the popular
``TS100'' soldering Iron. It implements an LCD and several push buttons
for adjusting temperature and other parameters. Internally, it runs a PID control
algorithm and an acceleration sensing routine for auto-power off.
Vulnerabilities in these devices can lead to overheating of the soldering iron,
which can cause damages to the objects being soldered or injuries to the
operators. 

{\em Heat Press}: This firmware corresponds to an industrial heat press
\cite{heat_press} used in a textile sublimation production line. The firmware
implements recipe manager for controlling the temperature, time and pressure of the
sublimation process. The system features a touch screen and a remote industrial
I/O channel using the Modbus protocol. Vulnerabilities in this type of systems
can lead to unintended operations, remote hijacking, and damage of industrial
facilities.

\point{Experiment Results}
Our framework achieved similar or even better results on real firmware than on the unit
tests. As shown in Table \ref{t-case-general-result}, the register
categorization accuracy (Acc.\%) is even higher than measured in the unit tests,
despite that the real firmware are more complex and access more registers and
peripherals. Our manual verification attributes this better result to the fact that these firmware follow the register access patterns more strictly than the sample 
drivers in the unit tests. We explain the reasons and impacts of register mis-categorizations in \S \ref{sec:miscat}. 
The last column of Table \ref{t-case-general-result} shows the total time (in
seconds) spent on model instantiation for each firmware. The time consumed (10
minutes in the worse case) is acceptable given that %
a typical fuzzing session
often lasts for days or longer. 

\begin{table}
    \centering
    \caption{Model instantiation statistics on 10 real firmware}
    \label{t-case-general-result}
    \begin{tabular}{@{}l@{}rrrrrr@{}}  
    \toprule
    \textbf{Firmware} &\textbf{Peri.}  &\textbf{Regs.}  &\textbf{Acc.} &\textbf{SR} &\textbf{Int.} & \textbf{Time}\\
    &&&\textbf{(\%)}&\textbf{group} &\textbf{line}&\textbf{(s)}\\
    \hline
    \textbf{Robot}	        &7	    &43	    &100.0	    &16	    &1 &131.2   \\
    \textbf{PLC}	        &5 	    &21		&100.0	    &5	    &2 &6.8     \\
    \textbf{Gateway}	    &14	    &101    &93.4	    &14		&5 &612.4   \\
    \textbf{Drone}          &11	    &68	    &100.0       &20     &2 &315.7   \\
    \textbf{CNC}            &12	    &81	    &91.5       &5      &2 &48.3    \\
    \textbf{Reflow O.}      &6	    &32	    &95.8       &1      &2 &4.4     \\
    \textbf{Console}        &11	    &43	    &88.5       &9      &1 &28.0    \\
    \textbf{Steering C.}    &11	    &79	    &69.6       &14     &3 &96.2    \\
    \textbf{Soldering I.}   &13	    &84	    &90.7       &33     &9 &512.0   \\
    \textbf{Heat Press}     &4	    &76	    &84.0       &25     &2 &59.4    \\
    \hline
    \end{tabular}
\end{table}

\begin{figure*}[t]
    \centering
    \includegraphics[width=17.5cm]{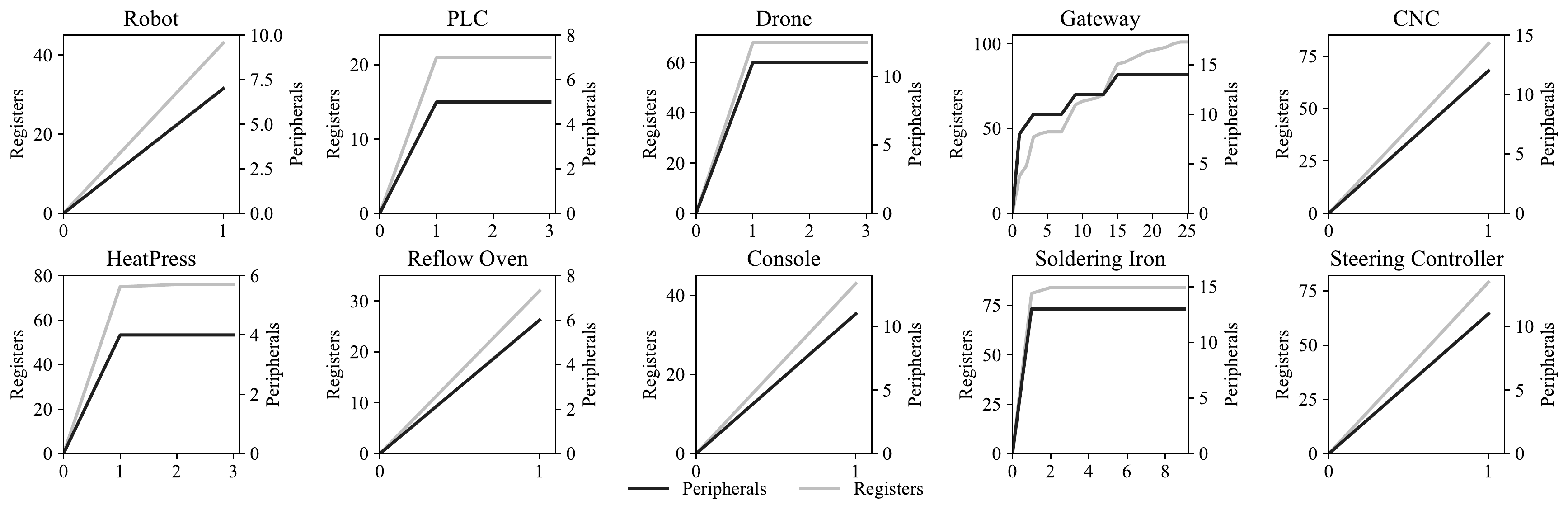}
    \caption{Per-round Progression of Model Instantiation (registers and peripherals covered) on 10 Real Firmware}
    \label{fig:FuzzingInstantiation}
\end{figure*}

Figure \ref{fig:FuzzingInstantiation} shows the progress of the model
instantiation for each firmware. As explained in \S \ref{sec:inst}, \sysname
launches a new round of model instantiation when it encounters an unmodeled or 
unhandled peripheral interface access. 
For most firmware, \sysname instantiated the models within 3 rounds. 
Note that the last few rounds of model instantiation for
PLC, Drone, HeatPress and Soldering Iron formed new SR read groups, which are
not shown in the figure for simplicity. Gateway incurs 25 rounds model
instantiation because it initializes peripherals on-demand (i.e., additional
model instantiation is needed when a new peripheral is initialized after the
model has stabilized). On the 10 tested firmware, \sysname triggers a round of 
model instantiation every 2,579 seconds, on average, 
until all peripheral interfaces are modeled. 

In most cases, \sysname automatically derived proper
models to support firmware execution with \propname satisfied (i.e., no crash,
stall, or operation skipping). We will present the two cases that break \propname 
in \S \ref{sec:miscat}. We then used these models to perform fuzzing on
all the firmware (\S\ref{sec:fuzzingcase}). 

This end-to-end experiment on real firmware and the unit tests 
firmware together confirm that our framework achieves its goal: enabling
hardware-independent and scalable firmware testing/fuzzing via \sysname.
Moreover, the overhead and inaccuracy are low enough for our framework to be
used in practice.

\subsection[]{Register Mis-categorizations \& False Crashes/Hangs}
\label{sec:miscat}
In this section, we discuss the two types of false positives, namely register mis-categorizations 
and false crashes/hangs, that our mechanism may cause.

\begin{table}
    \centering
    \caption{Numbers of mis-categorized registers on 10 real firmware, grouped by their impacts: either slowing down fuzzing (Type I) or limiting coverage (Type II). The last column shows the total number of registers that have been read during the firmware fuzzing process.  }
    \label{tab:miscat}
    \begin{tabular}{@{}l@{}c@{ }c@{ }c@{}}
    \toprule
    \textbf{Firmware}  & \textbf{Mis-cat. Regs } & \textbf{Mis-cat. Regs } & \textbf{Total Regs Read} \\
                       & \textbf{(Type I)} & \textbf{(Type II)} & \textbf{by Firmware} \\ 
    \hline
    \textbf{Robot}          &0  &0 &25 \\
    \textbf{PLC}            &0  &0 &18 \\
    \textbf{Gateway}        &4  &0 &61 \\
    \textbf{Drone}          &0  &0 &39 \\
    \textbf{CNC}            &1  &3 &47 \\
    \textbf{Reflow O.}      &0  &1 &24 \\
    \textbf{Console}        &0  &3 &26 \\
    \textbf{Steering C.}    &6  &1 &23 \\
    \textbf{Soldering I.}   &4  &1 &54 \\
    \textbf{Heat Press}     &4  &0 &25 \\
    \hline
    \textbf{Total \# (\%)}  &19 (5.6\%) &9 (2.6\%) &342 (100\%)  \\
    \hline
    \end{tabular}
\end{table}

\textbf{Register Mis-categorizations:} We manually examined all the registers mis-categorized by \sysname while being tested against the 10 real firmware. We  
itemized their potential impacts on firmware execution and present number of 
mis-categorized registers per impact in Table \ref{tab:miscat}. 
For each impact, we give 
a representative example of the mis-categorized register and 
explain why the register is mis-categorized. 

When calculating the register categorization accuracy, we only consider
registers that have been read at least once during firmware execution (i.e.,
registers never read are not counted because they do not affect the firmware
execution). In Table~\ref{tab:miscat}, the last column shows the total number of
registers that have been read by the firmware. The middle two
columns show the number of registers mis-categorized by \sysname for each
firmware. The mis-categorized registers are grouped in two types based on their
negative impact to firmware fuzzing (i.e., slowing down fuzzing or reducing the
coverage). The overall accuracy of register categorization (the ``Acc.'' column
in Table~\ref{t-case-general-result}) for each firmware is calculated as
follows: $Accuracy = 1 - (TypeI + TypeII) / TotalRegistersRead$.

Type-I mis-categorizations, mostly \sreg mis-categorized to \dreg, may slow
down the fuzzing process but do not stop or break it. 
The reason is that this type of mis-categorizations causes
the fuzzer to guess the \sreg value that is supposed to be quickly generated by
\sysname. Nonetheless, AFL can effectively guess the proper value using the
coverage information as guidance. %
Mis-categorized registers of this type are caused by certain peripheral drivers
not following the access patterns we defined to categorize registers. We
analyzed these cases manually and found that these drivers should have followed
the access patterns to avoid potential I/O errors. One exception is that the
firmware writes \sreg (to clear potential peripheral errors) before ever reading
it (to check the peripheral state), which causes the \sreg to be mis-categorized
to \dreg due to the write-on-first-access pattern for identifying \dreg.

Type-II mis-categorizations, mostly \dreg mis-categorized to \creg, prevent
the fuzzer from reaching some code paths that depend on the input from the
mis-categorized \dreg. Such
mis-categorized \dreg may stall part of the firmware (e.g., one thread of the
Soldering Iron firmware) and partially break \propname. This is because \sysname
is unable to channel the fuzzer-generated input into the firmware through the
mis-categorized register. 
The cause of such register mis-categorizations, similar to the previous type, is that drivers fail to follow 
the common/correct register access patterns to avoid 
potential I/O errors. One exception is that some GPIO peripherals expose  
multiple pins via one \dreg. To write data to a pin, a driver has to follow the RMW pattern 
to avoid overwriting other pins, which causes \dreg to be mis-categorized into \creg. 
This is a limitation of our register categorization method.

In summary, only 8.2\% registers read by firmware were mis-categorized, 
and less than one third (2.6\% out of 8.2\%) of the
mis-categorized registers negatively impacted \propname. This is empirically acceptable
because, as shown in \S\ref{sec:fuzzingcase}, \sysname achieves good fuzzing
performance on real firmware despite the (rare) mis-categorized registers.
\sysname enables hardware-free and scalable MCU firmware testing. It achieves
high code coverage and finds previously unknown bugs, with no false crashes and
very few false hangs.

\textbf{False Crashes/Hangs:}
In our evaluation, we found that no crashes/hangs are caused by infeasible 
input or code paths introduced by \sysname (\S \ref{sec:fp}). This justifies our 
design decision that leaves the task of input pruning to the fuzzer.
However, we found two hangs on the Soldering Iron firmware caused by the limitations 
of \sysname: one is caused by a \dreg that is mis-categorized as \creg (discussed 
above) and the other is due to firmware's usage of DMA, which is very difficult 
to model automatically and we consider it out of the scope for this paper. 

Except for the two hangs, all other crashes/hangs found by \sysname 
are caused by real bugs inside the firmware. 
We verified this by running the firmware on real devices with the 
same inputs that caused crashes/hangs on \sysname, and then confirmed that 
these inputs also cause crashes/hangs on the real devices.

\subsection[]{Fuzzing Results \& Case Studies}
\label{sec:fuzzingcase}

In this section, we demonstrate that \sysname, without requiring any MCU
hardware or peripherals, is able to fuzz-test real MCU firmware and find previously
unknown bugs. In our experiments, we used the unmodified AFL as the drop-in
fuzzer for \sysname\footnote{Any existing or future fuzzers can be used as a
drop-in component}. We did not use or evaluate other fuzzers because finding
or designing a better fuzzer is not the goal of this work. 
\sysname feeds AFL-generated inputs to firmware execution via \dreg (identified
during the model instantiation process) when accessed. 
\sysname collects and sends the execution coverage to AFL as the fuzzing feedback.

For simplicity, we used randomly generated seed inputs
(i.e., no expert knowledge about firmware input is given to the fuzzer). %
For testing purposes, our framework uses a very basic memory error detector, 
which is based on the segment tracking heuristics described in
\cite{wycinwyc}. It enforces the least permissions needed by each memory
segment: {\tt R+X} for flash, {\tt R+W} for RAM, peripherals, and system control
block, and {\tt no-access} for the rest of the memory. As a result, it can
only detect a small set of bugs (i.e., memory corruptions that span region 
boundaries and violate the permissions).

After fuzzing each firmware for 24 hours, our framework found 7 unique and 
previously unknown bugs in the firmware (Table \ref{t-case-bugs-result}). 
All of them were later confirmed as exploitable by our manual analysis. 
Based on the result, we can reasonably anticipate that our framework is likely 
to find even more bugs in these firmware by using expert-crafted seed input 
and an advanced MCU memory sanitizer (both are orthogonal to the topic of
this paper). We reported all the bugs to the device vendors.

\begin{table}[t]
    \centering
    \caption{Unique bugs detected in real firmware}
    \label{t-case-bugs-result}
    \begin{tabular}{@{}lll@{}}
    \toprule
    \textbf{Firmware} &\textbf{CWE*}  &\textbf{Unique bugs}   \\
    \hline
    \multirow{3}{*}{\textbf{PLC}}
    & 704, 129, 787       &3	   	     \\
    
    & 190, 129, 787       &1	   	     \\
    
    & 681, 129, 400       &1	   		 \\
    
    \hline
    \textbf{Gateway} & 129, 787           &1	\\		    
    \hline
    \textbf{Heat Press} & 129, 787         &1    \\          
    \hline
    \multicolumn{3}{l}{\scriptsize{* Common Weakness Enumeration (www.mitre.org)}} \\
    \end{tabular}
\end{table}

As shown in Table \ref{t-case-bugs-result}, the bug found in Gateway is a
combination of an improper validation of array index (CWE-129) and an
out-of-bound write (CWE-787). This bug allows a remote attacker to overwrite
data objects on the embedded device and cause denial of service or data
corruption. The bug found in Heat Press shares a very similar nature with 
the Gateway bug. The five bugs found in PLC, three similar and two distinct in
nature, are combinations of the common programming errors, such as incorrect
type cast (CWE-704), integer overflow (CWE-190), incorrect conversion between
numeric types (CWE-681), and uncontrolled resource consumption (CWE-400).

Some of these bugs are more critical than others due to the possibility of
arbitrary memory read/write by remote attackers. To demonstrate the real
security impact, we developed a proof-of-concept (PoC) for the PLC bugs.
As shown in Figure~\ref{fig:escenario}, the PLC device is typically attached to 
a fieldbus through Serial Port. Any malicious devices on the bus, either owned by an 
adversarial insider or compromised by a remote attacker, can exploit the PLC bugs
we found by sending crafted commands over the fieldbus. By the PoC, we
modified the internal memory arrays in PLC which contain the critical parameters
for PID (Proportional, Integral, Derivative control algorithm).
Using this PoC, an attacker can directly influence the PLC-controlled industrial
process, causing Stuxnet-like damages.

\begin{table}
    \centering
    \caption{Fuzzing performance on 10 real firmware}
    \label{t-fuzz-performance}
    \begin{tabular}{@{}lcccr@{}}
    \toprule
    \textbf{Firmware} &\textbf{Speed}   &\multicolumn{3}{c}{\textbf{Basic block coverage}} \\
                      &\textbf{(run/s)} &\textbf{w/o \sysname} & \textbf{w/ \sysname} & \textbf{Improv.} \\ 
    \hline
    \textbf{Robot}          &29.5   &2.5\%  &43.1\% &17.0x \\
    \textbf{PLC}            &32.7   &3.5\%  &26.1\% & 7.6x \\
    \textbf{Gateway}        &17.8   &1.7\%  &45.2\% &26.5x \\
    \textbf{Drone}          &17.2   &8.4\%  &58.4\% & 6.9x \\
    \textbf{CNC}            &18.0   &2.7\%  &69.5\% &26.1x \\
    \textbf{Reflow O.}      &24.7   &3.6\%  &39.8\% &11.2x \\
    \textbf{Console}        &14.6   &2.2\%  &37.8\% &17.2x \\
    \textbf{Steering C.}    &32.3   &0.7\%  &19.8\% &29.5x \\
    \textbf{Soldering I.}   &13.5   &4.2\%  &53.2\% &12.7x \\
    \textbf{Heat Press}     &39.4   &1.1\%  &28.1\% &24.8x \\
    \hline
    \end{tabular}
\end{table}

This fuzzing experiment not only demonstrates our framework's ability to find
bugs in real firmware but also shows its relatively high level of code coverage
and fuzzing speed. Table
\ref{t-fuzz-performance} shows the fuzzing speed (number of fuzzing runs per
second), the basic block coverage without \sysname, the basic block coverage
with \sysname, and how much coverage \sysname improves. 
With \sysname, the code coverage improved 7 to 30 times from the
coverage without \sysname, echoing the value of \sysname and the importance of
(automatic) peripheral I/O handling. 

The much-improved code coverage may still
seem low number-wise. We investigated it and found four main causes for it.
First, these firmware tend to contain dead code as regular software does (i.e.,
unused or fractionally used libraries). Second, AFL is a simple grey-box fuzzer
that is not good at 
breaking through
complex path conditions (e.g., checksum checks). Such inputs and path
conditions are commonly seen in firmware. %
Since \sysname hosts fuzzer as a drop-in component, we can replace AFL with
a more advanced fuzzer such as \cite{taintscope} to overcome this problem. 
Third, the two false hangs on Soldering Iron firmware causes fuzzer unable to 
cover part of the firmware. 
Fourth, we identified that not only the input values, 
but also the input duration (i.e., how long an input value/signal maintains), affect firmware 
execution on embedded devices. However, neither \sysname nor any existing fuzzer considers 
input duration, which poses an open challenge for future research. For example, Soldering Iron 
and Reflow Oven firmware constantly read from GPIO while performing different 
operations determined by the duration of the same GPIO value/signal. 
Existing fuzzers generate inputs without considering input durations. Despite
the existence of a functioning timer in \sysname (by which the firmware can
measure the duration of a GPIO value), firmware operations triggered by long
durations of GPIO values can not be executed. The lack of support for varying
input duration, although can be partly mitigated by \sysname, needs to be better
addressed by fuzzers, which are the dedicated component for input generation. We
believe that 
input duration is a unique challenge for MCU firmware fuzzing, which is out of
the scope of this paper and is an interesting topic to study for future works.
We also identified two reasons as to why we could not detect even more bugs than
the 7 reported in Table \ref{t-case-bugs-result}. First, some firmware designs
are not susceptible to memory corruption errors. For example, the Reflow Oven
and Steering Control firmware rarely use buffers or dynamically allocated memory
for performance reasons. Therefore, such firmware is unlikely to have memory
corruption bugs. Second, the bare minimum memory error detector (or sanitizer)
that we implemented may fail to catch the bugs triggered by the fuzzer. For
example, we manually identified 1 extra bug on the PLC firmware. The bug,
although similar to those found during fuzzing (Table \ref{t-case-bugs-result}),
was triggered by \sysname during testing but went undetected by the simple
memory error detector.
\section{Discussion}
\label{sec:discuss}

\point{Direct Memory Access (DMA)}
\sysname models the processor-peripheral interfaces, including registers and
interrupts. It does not model Direct Memory Access (DMA), which allows
peripherals to directly access RAM and in turn provide input to firmware. The
lack of DMA support is a limitation of our work. Due to DMA's complex and
peripheral-specific nature, modeling DMA is arguably impossible without
considering internal peripheral designs, which goes against \sysname's design
principle---being generally applicable to a wide range of peripherals and MCU
devices. Nonetheless, the usage of DMA depends on the design and 
architecture of individual firmware. We observed that most of the MCUs studied and tested 
support DMA. However, only 1 out of the 10 real firmware tested in \S\ref{sec:firmwaretest} actually uses DMA. 

\point{Architectures beyond ARM}
We analyzed 3 MCUs that use non-ARM architectures for IoT devices:
ATmega328P (AVR), PIC32MX440F256H (MIPS) and FE310-G000 (RISC-V). 
Our analysis shows that our design of \sysname 
and the abstract model that we defined are not specific to the ARM architecture. They can be extended to support the other architectures such as AVR, MIPS and RISC-V. All these non-ARM architectures define specific 
memory-mapped areas for peripherals similar to ARM. They also follow similar register 
categories (\creg, \sreg, \dreg and \csreg) that \sysname identifies. Furthermore, the procedures to configure and 
operate peripherals on these non-ARM architectures follow the same conventions and patterns 
that \sysname uses for recognizing and handling peripheral I/O.

We observed a slight difference in accessing memory-mapped I/O by AVR. AVR can 
use either specific opcodes (IN/OUT) or ST/LD instructions 
to access mapped peripheral registers. ST and LD instructions require a constant
offset to access the same addresses accessed by IN/OUT opcodes. 
We also observed that RISC-V implements a unique interrupt handling mechanism 
that uses Hardware Threads (HART) and a Platform-Level Interrupt Controller (PLIC). 
RISC-V also uses a new type of hybrid register ({\tt S\&DR}) that is not seen on the other architectures. 
\sysname and the current abstract model can be extended to handle  
these architectural differences and in turn support these non-ARM MCU architectures.

\point{Firmware Analysis beyond Fuzzing}
Although our work was initially inspired by the open challenges facing firmware
fuzzing, it is not designed to support fuzzing exclusively. Other types of
dynamic firmware analysis that do not require fully accurate output from
firmware can use our framework to achieve hardware-independence and scalability.
For instance, data or code reachability analysis, such as taint analysis and
certain debugging tasks, can benefit from our framework. In particular, concolic
firmware execution can use our framework to generate more realistic concrete
inputs (i.e., non-crashing/stalling), reduce the number of symbolic values, and
avoid some infeasible code paths.  %
\section{Related Work}
\label{sec:related}

\point{Dynamic Firmware Analysis}
Several recent works addressed the high barrier of dynamically analyzing MCU
firmware. They follow the hybrid emulation approach, which forwards peripheral
operations to real devices while running  firmware on a customized emulator. 
Avatar \cite{avatar} proposed a novel framework for hybrid emulation and used it
for conducting concolic execution \cite{s2e}. Surrogates \cite{surrogates}
significantly improves the forwarding performance of Avatar via customized
hardware. A follow-up work \cite{wycinwyc} fuzz-tested simple programs with
manually injected vulnerabilities using Avatar and revealed that, without an
effective sanitizer for MCU, fuzzers by themselves cannot observe many bugs even
if they are triggered. Avatar2 \cite{avatar2} extends Avatar to allow replay of
forwarded peripheral I/O without using real devices. Charm \cite{charm} targets
smartphone drivers, rather than MCU firmware. It adopts the forwarding approach
similar to Avatar. Prospect \cite{prospect} forwards peripheral accesses at the
syscall level, which however does not exist on bare-metal MCU devices.
\cite{peri_caching} uses cached peripheral accesses to approximate firmware
states for analysis. 

These works collectively improved the state of the art of dynamic firmware
analysis. However, they have heavy hardware dependence, which is at odds with
speedy and scalable fuzzing. Plus, they require significant expert-knowledge and
human efforts to set up and run. In contrast, our framework is largely automated
and completely removes hardware dependency from dynamic firmware analysis,
without using peripheral I/O forwarding or replaying. Moreover, we found bugs
from real MCU firmware whereas none of the previous works did, echoing the value
of the scalable fuzzing enabled by our framework.

Another line of works analyzed firmware running in fully emulated
environments~\cite{firmadyne,dyn_webif,symdrive} or directly on the target
hardware~\cite{periscope}. Instead of MCU devices, these works target
Linux-based devices, which are closer to general-purpose computers than truly
embedded devices. Linux-based devices  have  much better emulator support due to
the much less diverse peripherals than MCU devices. Analyzing firmware of these
devices does not face the MCU-specific challenges that our work overcame.

\point{Static Firmware Analysis}
FIE \cite{fie} applied symbolic execution to TI MSP430 firmware by extending
KLEE \cite{klee} with a peripheral model. It returns an unconstrained symbolic
value for each peripheral register read. It assumes any enabled interrupts can
happen after every instruction, which caused the state explosion problem.
Inception \cite{inception} address this problem by optionally forwarding
peripheral access to hardware using Avatar \cite{avatar}.
FirmUSB \cite{firmusb} proposed a symbolic execution mechanism tailored for USB
controller firmware on 8051/52 architectures using domain specific knowledge.
Firmalice \cite{firmalice} aims to find authentication bypass vulnerabilities in firmware
via concolic execution. A large-scale study on Linux-based firmware
\cite{stat_linux} reported presence of weak passwords and known-vulnerable code.
Although addressing the same high-level problem of firmware security, these
works follow an orthogonal approach from ours (static vs. dynamic analysis). 

\section{Conclusion}
\label{sec:conc}

We presented \sysname, a novel technique for modeling the I/O behaviors of the
processor-peripheral interfaces. It is the first to enable peripheral-oblivious
emulation of MCU devices, and in turn, allow MCU firmware to be dynamically
tested with high code coverage, at scale, and without hardware dependence. We built
\sysname into a framework that executes a given firmware binary and hosts a
drop-in fuzzer (AFL) as the input source. We evaluated the framework using 70
sample firmware and 10 real device firmware. It fully booted and tested 79\% of
the firmware without any human intervention. When paired with a limited memory
error detector, it found 7 new bugs from the real device firmware. The results
suggest that our framework is of great value and potential for practical use. 

\bibliographystyle{plain}
\bibliography{refs}

\appendix

\section[]{Vulnerability Analysis and PoC: PLC firmware}
\label{sec:poc}
We confirmed all 7 bugs found on PLC, Gateway and Heat Press firmware (Table 
\ref{t-case-bugs-result}) are exploitable by Proof of Concepts (PoC). For 
simplicity, we present PoC for one bug found on PLC firmware. 

\begin{figure}
    \centering
    \begin{subfigure}{0.22\textwidth}
      \centering
      \includegraphics[width=3.5 cm]{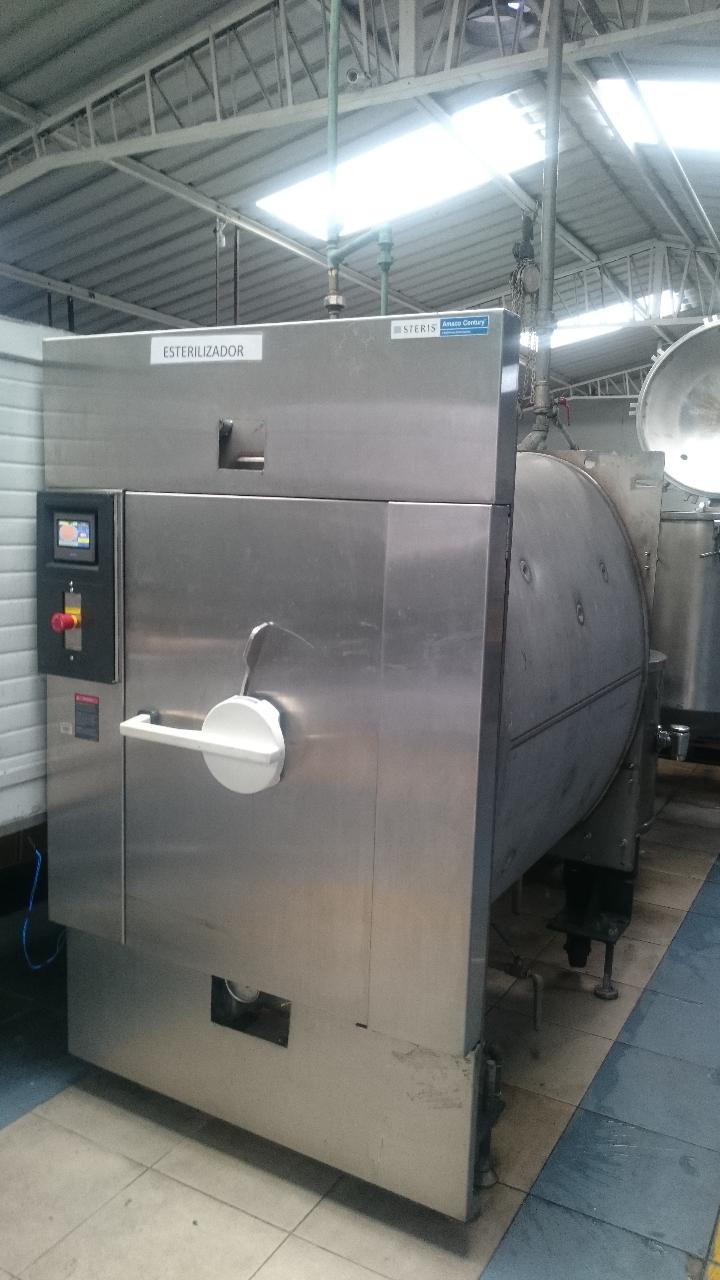}
      \caption{Front panel}
      \label{fig:sterilyzerA}
    \end{subfigure}%
    \begin{subfigure}{0.22\textwidth}
      \centering
      \includegraphics[width=3.5cm]{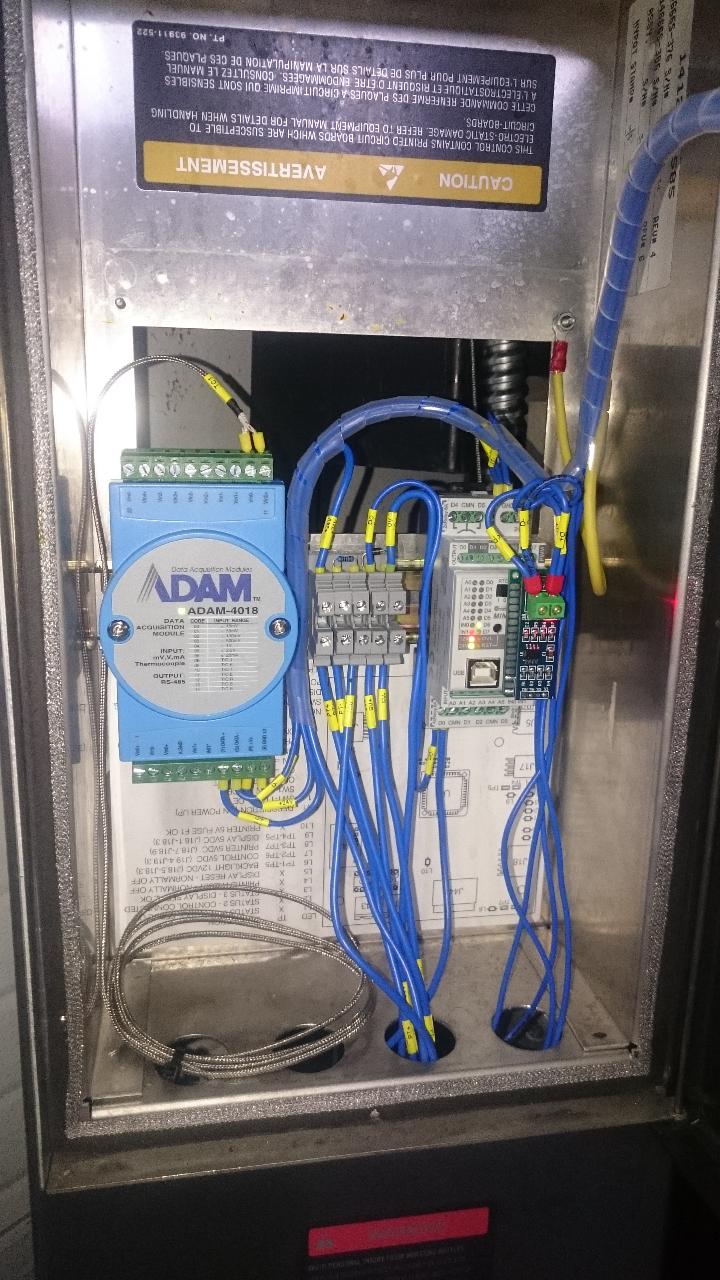}
      \caption{PLC control panel}
      \label{fig:sterilyzerB}
    \end{subfigure}
    \caption{Sterilizer used in an industrial food processing facility, whose firmware was among the 10 real firmware we tested}
    \label{fig:sterelyzer}
\end{figure}

The PLC firmware tested was obtained from a sterilizer machine in an
industrial food facility (figure \ref{fig:sterelyzer}). This firmware contains
the standard OS functionality of Arduino, the user/control logic (PID loops,
schedule), and the communication stack. The communication stack implements the Modbus
RTU slave protocol. This protocol uses USART peripheral over the RS485 standard
to access the fieldbus. The PLC receives Modbus queries from a human-machine 
interface (HMI) or SCADA, which is also connected to the fieldbus.

The OS and the communication stack are based on the Arduino framework with some
modifications made by the equipment vendor. We obtained the source code of the
communication stack (Modbus) from  the  public repository of the hardware
vendor. We ported this AVR source code to Cortex-M without any manual changes
thanks to Arduino compatibility. This is the firmware that we used on our
fuzzing session.

Our analysis determined that the crashes triggered during the fuzzing session
result from a buffer overflow bug in the Modbus implementation. The root
cause of this bug is a wrong casting in the validation of Function Code 15
(FC15) of the Modbus protocol. This function code is used to force multiple
coils (write multiple memory locations). The field that controls the number of
coils can be crafted to bypass the validation of FC15. This bug allows attackers
to overwrite device memory.

The Vulnerability that we discovered can be used to launch Stuxnet-like attacks
remotely. The reason is that FC15 provides attacker two primitives to control
where and what to write on PLC memory. We verified that, although the memory
span controlled by the attacker is limited, it is still possible to overwrite
important PLC data structures. We implemented a POC to demonstrate this capability.

\begin{figure}
    \includegraphics[width=6cm]{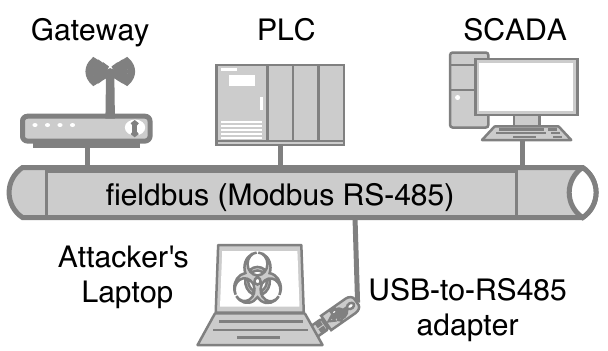}
    \caption{Scenario of a remote attack on Modbus over RS485}
    \label{fig:escenario}
\end{figure}

To run the POC, we used the NUCLEO144-F429ZI development board to play the role
of a PLC running the vulnerable firmware. We connected the USART of our PLC to
our laptop (i.e., attacker) trough a USB-to-Serial adapter. We modified an open-source
version of the Master Modbus library to craft a FC15 Modbus master query,
which runs in the attacker laptop. In addition, we connected a computer to the
debugging port of the PLC to check the result of our attack. This setup
implements the attack scenario presented in figure \ref{fig:escenario}. 

\begin{figure}
    \includegraphics[width=\linewidth]{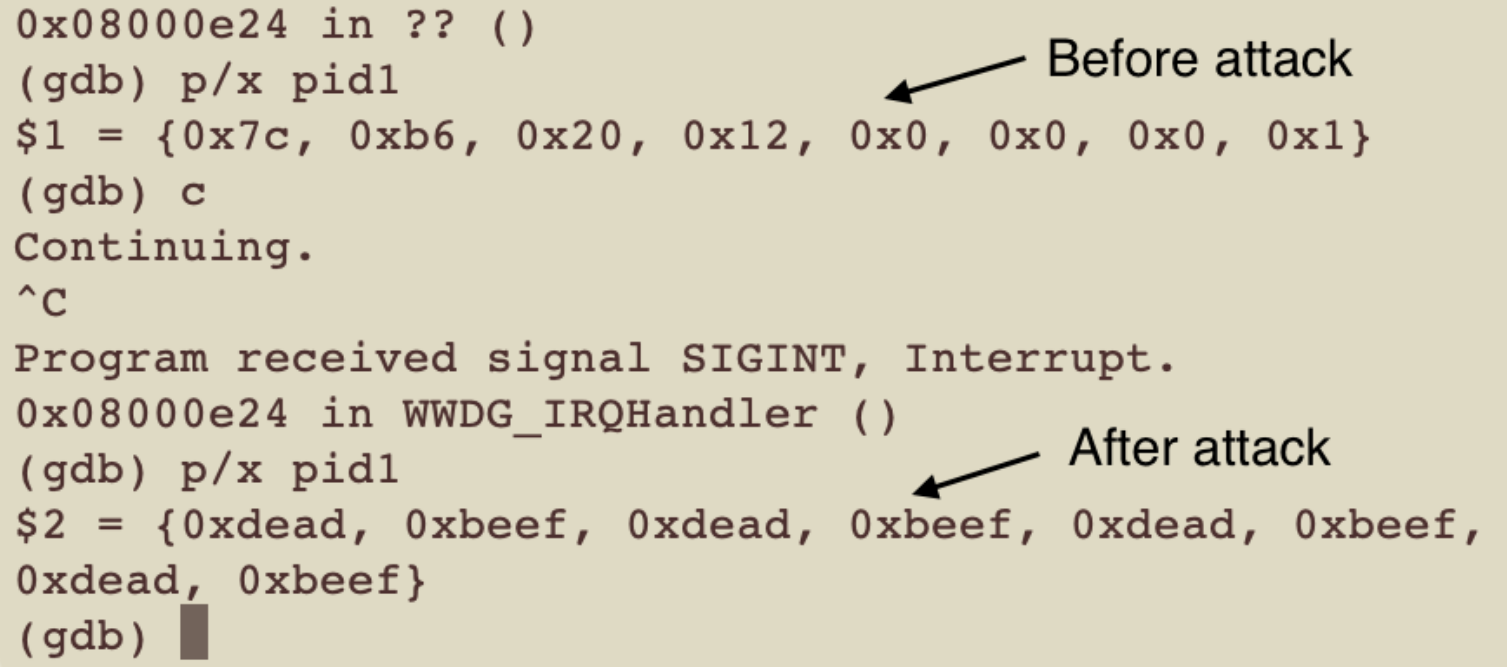}
    \caption{PID data structure before and after attack; It is overwritten to 0xdead 0xbeef.}
    \label{fig:attack}
  \end{figure}

Figure \ref{fig:attack} shows the output of the debugging console of the PLC and
confirms the success of the POC (i.e., memory being overwritten with an
attacker-supplied value). It demonstrates that our framework was able to
identify bugs with security consequences on real production firmware.
Consequently, \sysname is able to model the peripheral-processor interface of
USART and automatically channel data from the fuzzer engine to test upper layers
(Modbus) of real firmware.

\newpage
\section{Full Results of Unit Tests}
\label{AppendixGeneralResult}
Table \ref{t-AppendixGeneralResult} shows the full results for all the 70 unit
tests (\S\ref{sec:unittest}). We wrote ``N/A'' for the 14 tests where the combinations of MCU Soc and
OS/libraries are impossible due to incompatibility.

\vfill\break

\begin{sidewaystable}
\centering
\caption{General results of evaluation of unit test}
\label{t-AppendixGeneralResult}

\begin{tabular}{|m{1.4cm}|m{3.9cm}|m{0.9cm}|m{0.8cm}|m{0.9cm}|m{0.8cm}|m{0.9cm}|m{0.8cm}|m{0.9cm}|m{0.8cm}|m{0.9cm}|m{0.8cm}|m{0.9cm}|m{0.8cm}|}
\hline
\multicolumn{2}{|c|}{  }                                                              &   \multicolumn{2}{l}{\textbf{F103/Arduino}}	&	\multicolumn{2}{|c|}{\textbf{F103/NUTTX}}		&	\multicolumn{2}{|c|}{\textbf{F103/RIOT}}   &	\multicolumn{2}{|c|}{\textbf{K64F/RIOT}}	   &	\multicolumn{2}{|c|}{\textbf{SAM3X/Ardu.}}		&	\multicolumn{2}{|c|}{\textbf{SAM3X/RIOT}}    \\   
\hline

\textbf{Peripheral}  &\textbf{Functional Operation}	   &\textbf{Result}	&\textbf{Acc.\%}	            &\textbf{Result}	&\textbf{Acc.\%} &\textbf{Result} &\textbf{Acc.\%}	 &\textbf{Result} &\textbf{Acc.\%}  &\textbf{Result} &\textbf{Acc.\%}	 &\textbf{Result} &\textbf{Acc.\%}    \\
\hline
\multirow{2}{1cm}{\textbf{SPI}}                 	    & Receive a byte	                    &	Pass	&	94.74	                &	Pass	&	83.33	                &	Pass	&	83.33	            &	Pass	&	93.55	               &	Pass	&	77.27	                    &	Pass	&	100                  \\
           	             &Transmit a byte	                    &	Pass	&	94.74	                &	Pass	&	83.33	                &	Pass	&	83.33	            &	Pass	&	93.55	               &	Pass	&	77.27	                    &	Pass	&	100                  \\
\hline         

\multirow{2}{1cm}{\textbf{USART}}         	    & Receive a byte	                    &	Pass	&	100	                &	Fail	&	91.67	                &	Fail	&	71.43	            &	Pass	&	92.86	               &	Pass	&	75.00	                    &	Pass	&	100                  \\
	                    & Transmit a byte	                    &	Fail	&	100	                &	Pass	&	91.67	                &	Pass	&	71.43	            &	Pass	&	92.86	               &	Pass	&	75.00	                    &	Pass	&	100                  \\

\hline   
\multirow{2}{1cm}{\textbf{I2C}}                 	    & Read a byte from slave	                    &	Pass	&	100	                &	Fail	&	85.00	                &	Pass	&	75.00	            &	Fail	&	89.66	               &	Fail	&	78.95	                    &	N/A	    &	N/A                     \\
	                        & Write a byte to slave	                    &	Pass	&	100	                &	Fail	&	85.00	                &	Pass	&	75.00	            &	Pass	&	89.66	               &	Fail	&	78.95	                    &	N/A	    &	N/A                     \\
\hline

\multirow{3}{1cm}{\textbf{GPIO} }              	    & Exec. callback after pin int. 	&	Fail	&	85.71	                &	Pass	&	89.47	                &	Fail	&	82.35	            &	Fail	&	85.29	               &	Fail	&	73.91	                    &	Fail	&	75.00                   \\
  	                    & Read status of a pin	                                &	Fail	&	85.71	                &	Pass	&	84.21	                &	Pass	&	81.82	            &	Pass	&	85.29	               &	Pass	&	77.78	                    &	Pass	&	100                  \\
                       & Set/Clear a pin	                                    &	Pass	&	85.71	                &	Pass    &	84.21	                &	Pass	&	81.82	            &	Pass	&	85.29	               &	Pass	&	77.78	                    &	Pass	&	77.78                   \\
\hline

\textbf{ADC}	                        & Read an analog-to-digital conversion   &	Pass	&	96.67	                &	Pass	&	95.24	                &	N/A	    &	N/A	                &	Pass	&	64.44	               &	Pass	&	69.57	                    &	Pass	&	100                  \\
\hline
\textbf{DAC}                        & Write a value for digital-to-analog conversion	    &	N/A	    &	N/A	                    &	N/A	    &	N/A	                    &	N/A	    &	N/A	                &	N/A	    &	N/A	                   &	Pass	&	69.57	                    &	Pass	&	100                  \\
\hline

\multirow{2}{1cm}{\textbf{TIMER}}               	    & Exec. callback after int.	    &	N/A	    &	N/A	                    &	N/A	    &	N/A	                    &	Fail	& 	75.00	            &	Pass	&	92.86	               &	N/A	    &	N/A	                        &	Pass	&	87.50                   \\
                      & Read Counter value	                            &	N/A	    &	N/A	                    &	N/A	    &	N/A	                    &	Pass	&	75.00	            &	Pass	&	92.86	               &	N/A	    &	N/A	                        &	Pass	&	87.50                   \\

\hline

\textbf{PWM}                 	    & Configure PWM as Auton.	&	Pass	&	94.44	                &	Pass	&	78.95	                &	N/A	    &	N/A	                &	Pass	&	86.21	               &	Pass	&	69.57	                    &	Pass	&	100                  \\

\hline

\multicolumn{2}{|c|}{\textbf{Accumulated results}} & Pass:8 Fail:3	&	94.34 &	Pass:8 Fail:3	    &	86.55	                &	Pass:8 Fail:3	    &	77.77	            &	Pass:11 Fail:2	    &	88.03	               &	Pass:9 Fail:3	    &	75.05	            &	Pass:11 Fail:1	    &	93.98                   \\
\hline
\end{tabular}
\end{sidewaystable}

\section{Firmware Information}
\label{sec:fm_info}
In this section, we extend the information for the real firmware tested in the end-to-end test (\S\ref{sec:firmwaretest}) 
and fuzzing (\S\ref{sec:fuzzingcase}). Table \ref{t-case-study-firmware} presents, for each firmware, its 
target MCU, OS, Lines of Code (LoC), size, source, and picture of product. The reported size of firmware 
corresponds to the compiled ELF file, which may contain extra information such as headers, 
sections and debug symbols.

  \begin{table*}[t]
    \centering
    \caption{Real Firmware Tested}

    \label{t-case-study-firmware}
    \begin{tabular}{m{1.6cm} m{2.8cm}  m{1.8cm}  m{2.0cm}  m{2.5cm} m{3.0cm }} \\ 
    \toprule
    \textbf{Firmware} & \textbf{MCU} & \textbf{OS/Sys lib.} & \textbf{LoC/size} &  \textbf{Source/Product} &\textbf{Product Image} \\
    \midrule
    \textbf{Robot}            & STM32F103RB     & Bare metal   & 32,999/960KB & \cite{robot_source} / DIY     & \begin{minipage}[c]{.3\textwidth} \includegraphics[ width=25mm]{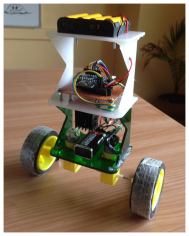} \end{minipage}\\[0.15cm]
    \textbf{PLC}              & STM32F429ZI     & Arduino      & 10,578/774KB & \cite{ControllinoModbusLibray}* / \cite{ControllinoMaxi}      & \begin{minipage}[c]{.3\textwidth} \includegraphics[ width=25mm]{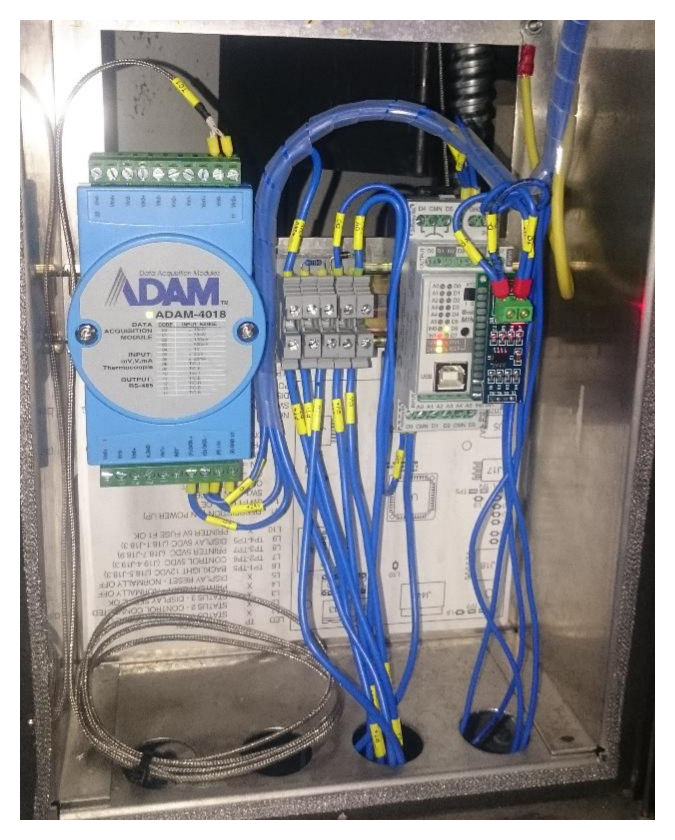} \end{minipage}\\[0.15cm]
    \textbf{Gateway}          & STM32F103RB     & Arduino      & 12,655/917KB & \cite{firmata} / DIY                                                        & DIY  \\[0.15cm]
    \textbf{Drone}            & STM32F103RB     & Bare metal   & 11,163/425KB & \cite{drone_source}   / \cite{PlutoDrone}               & \begin{minipage}[c]{.3\textwidth} \includegraphics[ width=25mm]{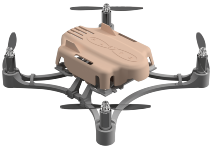} \end{minipage}\\[0.15cm]
    \textbf{CNC}              & STM32F429ZI	    & Bare metal   & 7,561/287KB & \cite{cnc_source}     / \cite{cnc_product}              & \begin{minipage}[c]{.3\textwidth} \includegraphics[ width=25mm]{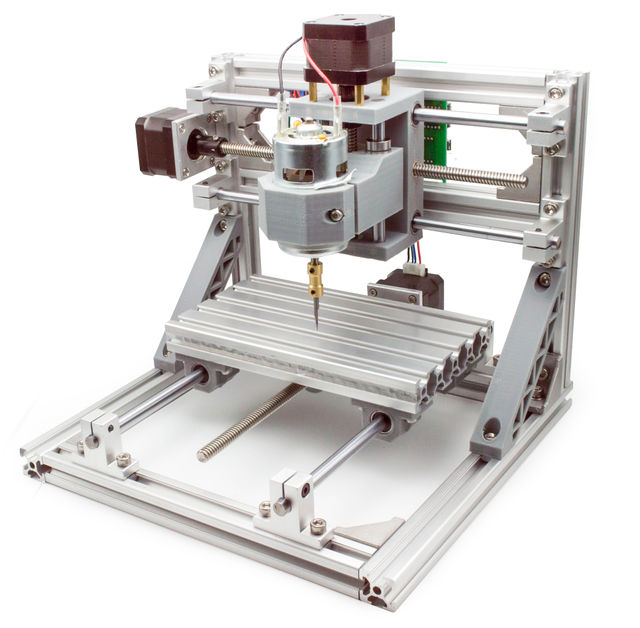} \end{minipage}\\[0.15cm]
    \textbf{Reflow Oven}      & STM32F103RB     & Arduino      & 12,272/820KB & \cite{reflow_source}  / \cite{reflow_product}           & \begin{minipage}[c]{.3\textwidth} \includegraphics[ width=25mm]{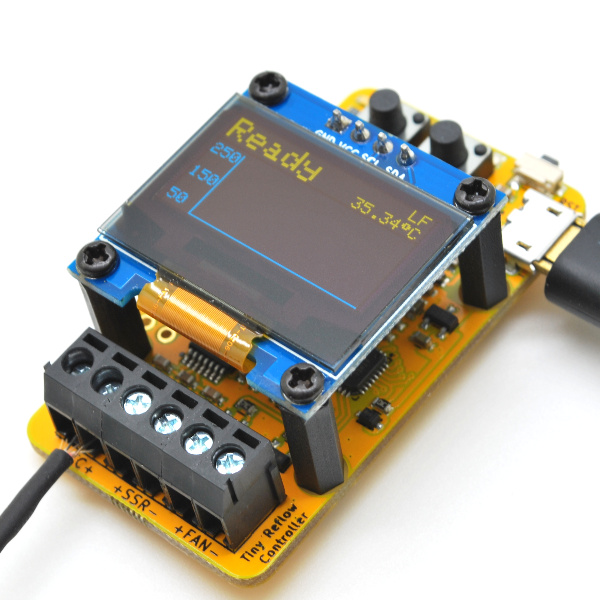} \end{minipage}\\[0.15cm]
    \textbf{Console}          & MK64FN1M0VLL12	& RIOT	       & 6,984/1,132KB & \cite{RiotOS} / DIY                                     & DIY   \\[0.15cm]
    \textbf{Steering Control} & SAM3X8E	        & Arduino      & 4,749/276KB & \cite{steering_source} / DIY                            & DIY   \\[0.15cm]
    \textbf{Soldering Iron}   & STM32F103RB	    & FreeRTOS	   & 43,928/491KB & \cite{soldering_source} / \cite{product_F103}           & \begin{minipage}[c]{.3\textwidth} \includegraphics[ width=25mm]{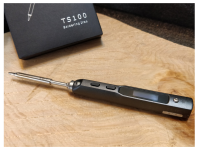} \end{minipage}\\[0.15cm]
    \textbf{Heat Press}       & SAM3X8E	        & Arduino      & 4,150/248KB & \cite{ControllinoModbusLibray}*/ Proprietary  & \begin{minipage}[c]{.3\textwidth} \includegraphics[ width=25mm]{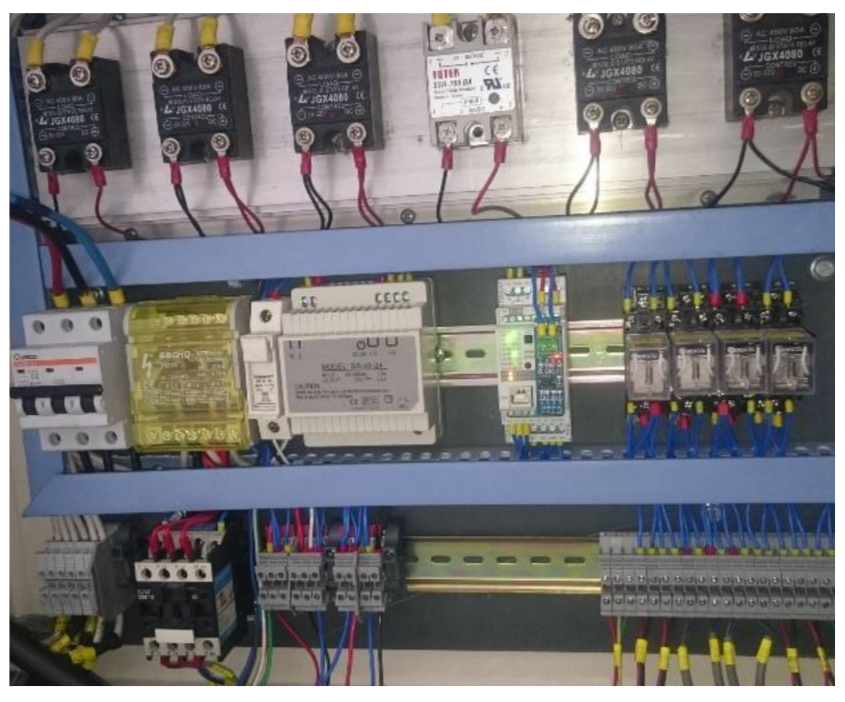} \end{minipage}\\
    \hline
    \multicolumn{6}{l}{*Only open-source libraries are disclosed, PLC/Machine control routines are property of their respective owners.} \\
    \bottomrule
    \end{tabular}
    
  \end{table*}
\end{document}